\documentclass[11pt]{article}

\usepackage{amsmath}
\usepackage[paper=a4paper,text={16cm,24cm},centering]{geometry}
\usepackage{graphicx}
\usepackage{ulem}
\usepackage{color}
\def\ignore#1{ }
\def\be{\begin{equation}}
\def\ee{\end{equation}}
\def\ruv{r_\text{\tiny UV}}
\def\Rir{{\mathcal R}}%_\text{\tiny IR}}
\def\uvec#1{\underline{#1}}
\DeclareMathOperator{\cotan}{cotan}
\def\rmax{r_{\text{max}}}

\begin{document}

\title{Gluon density fluctuations in dilute hadrons}

\author{Laura Domin\'e${}^{\footnotesize(1)}$, Giuliano Giacalone${}^{\footnotesize(2)}$,
  C\'edric Lorc\'e${}^{\footnotesize(3)}$, \\St\'ephane Munier${}^{\footnotesize(3)}$,
  Simon Pekar${}^{\footnotesize(4)}$\\
  \\
\footnotesize\it    (1)  Department of Physics, Stanford University, Stanford, CA 94305, USA\\
\footnotesize\it    (2) Institut de physique th\'eorique, Universit\'e Paris-Saclay, CNRS, CEA,
  91191 Gif-sur-Yvette, France\\
\footnotesize\it  (3) CPHT, \'Ecole Polytechnique, CNRS, Universit\'e Paris-Saclay, Route de Saclay, 91128 Palaiseau, France\\
\footnotesize\it  (4) ETH Z\"urich, Department of Physics, Wolfgang-Pauli-Strasse 27, 8093 Z\"urich, Switzerland}

\date{October 12, 2018}

\maketitle

\begin{abstract}
  Motivated by the relation existing between the gluon density in a hadron
  and the multiplicity of the particles measured in the final state of
  hadron-nucleus collisions, we study systematically the fluctuations of the gluon
  density in onia, which are the simplest dilute hadrons,
  of different sizes and at various rapidities.
  We argue that the small and the large-multiplicity tails of the
  gluon distributions present universal features, which should translate into
  properties of the multiplicity of the particles
  measured in the final state of high-energy proton-nucleus collisions,
  or of deep-inelastic scattering at a future electron-ion collider.
  We propose simple physical pictures of the rare events populating the tails of the
  multiplicity distribution
  that allow us to derive analytical formulas describing these universal behaviors,
  and we compare them to the results of Monte Carlo simulations.
\end{abstract}

\section{Introduction}

Measurements of the number of particles produced in the final states of high-energy hadron-nucleus (p-A) collisions have opened new windows in the study of dense partonic systems.
Indeed, the multiplicity of particles detected in p-A collisions in the region of fragmentation of the hadrons is expected to be sensitive to the properties of their partonic content at the time of interaction.
Nowadays, distributions of particle multiplicities are measured with great accuracy at particle colliders, most notably, in p-Pb collisions at the Large Hadron Collider (LHC)\cite{Adam:2014qja,Aad:2014lta,Aaij:2015qcq}, and the possibility of relating the experimental data to the small-$x$ dynamics of the parton distributions has triggered a lot of activity in the theoretical community~\cite{Dumitru:2011wq,Schenke:2012fw,Dusling:2018hsg}. 
A striking observation made in experiment is that multiplicity distributions in p-Pb collisions present a high-multiplicity tail
that is much longer than in Pb-Pb collisions, as expected from the fact
that p-A collisions are more fluctuation-dominated~\cite{Rogly:2018ddx}.
This tail may, then, contain valuable information about the wave functions of the projectile protons, and their event-by-event fluctuations.

Models of multiplicity fluctuations in the framework of high-energy hadronic collisions are mainly of two kinds.
On the one hand, we have purely phenomenological models, that serve as initial conditions for hydrodynamic calculations, and that are typically based on rather \textit{ad hoc} modifications of the Glauber model~\cite{Bozek:2013uha,Kozlov:2014fqa}.
These models have largely grown in complexity over the past few years, and include now prescriptions to model the sub-nucleonic degrees of freedom in the proton~\cite{Loizides:2016djv,Welsh:2016siu,Albacete:2016gxu,Moreland:2018gsh}.
Although such prescriptions are very successful in reproducing the experimental data, providing insight about the underlying dynamics of the parton distributions is beyond their scope.
A different kind of calculations of particle multiplicities, that takes as input the configurations of gluons inside protons and nuclei, have instead been achieved within the color glass condensate picture
\cite{Balitsky:1995ub,JalilianMarian:1996xn,JalilianMarian:1997gr,Iancu:2001md,Weigert:2000gi}
of high-energy quantum chromodynamics (QCD).
These calculations are also very successful in phenomenological applications.
A famous example is that of the number of gluons produced from the decay of color flux tubes in the glasma framework~\cite{Gelis:2009wh}, which is distributed according to a negative binomial distribution, whose long tail at large multiplicity provides naturally a good description of proton-proton data.

In this paper, we work in the theoretical framework of high-energy QCD, and, following the picture of particle production introduced in Refs.~\cite{Mueller:2016xti,Liou:2016mfr}, we argue that the multiplicity of particles probed around some off-forward rapidity in the region of fragmentation of the protons reflects, in each event, the integrated gluon density in the corresponding realization of the Fock state of the hadron at the time of interaction. 
In this picture, then, fluctuations of the multiplicity of particles are strictly related to the event-by-event fluctuations of the gluon density.
Instead of studying this phenomenon directly in the case of proton-nucleus collisions at the LHC, which will require to introduce some amount of modelization for the evolution of the proton, we focus on the simpler case of onium-nucleus collisions, where one can gain a very solid theoretical understanding in controlled asymptotic limits that allow to study multiplicity fluctuations analytically.
To this purpose, we shall work within the color dipole model (supplemented with an infrared cutoff for parton confinement), whose formulation is particularly well-suited to address this problem.

Our starting point is the main result of Ref.~\cite{Liou:2016mfr}, namely, an analytical estimate of the behavior of the high-multiplicity tail of the gluon number density in a boosted onium.
After reviewing this calculation, we extend the analysis to the opposite case of low-multiplicity events, and we derive a new formula for the behavior of the low-multiplicity region of the gluon number density.
However, both our calculation and that of Ref.~\cite{Liou:2016mfr} rely largely on conjectures, and leave important parameters undetermined.
The main thrust of this paper is, eventually, that of checking the validity of these derivations, and better understand them, by means of extensive Monte Carlo simulations of the small-$x$ evolution of the Fock state of an onium in the dipole picture.
Our goal is that of showing that the asymptotics of gluon density fluctuations in an onium present robust universal features. 

Our motivation for pursuing theoretical studies of multiplicity fluctuations in the dipole model, pioneered over
20~years ago by Salam~\cite{Salam:1995zd}, is twofold.
First, the theoretical understanding of dipole evolution has improved
since then, as well as the numerical capabilities, making possible much more accurate
evaluations of distributions. Second, and more importantly,
there is now a strong motivation to
better understand this physics, since the LHC is taking data for which such
studies are relevant and timely.
Our work may also be of interest for a future electron-ion collider,
and actually, easier to connect to the experimental data in that case:
Indeed, in a high-energy
scattering, the interaction
between an electron and an ion is mediated by a photon, whose $q\bar q$ (onium) component of the wave function is perfectly determined in the framework of quantum electrodynamics (QED).

Our paper is organized as follows.
In Sec.~\ref{sec:model}, we recall the connection between
the final-state multiplicity in hadron-nucleus collisions and the integrated gluon number density
in the hadron, and we explain how to compute the fluctuations of the latter in the case in which
the hadron is a heavy onium.
In Sec.~\ref{sec:asymptotics},we propose physical pictures of the events populating the tails of the gluon number distributions, and we establish analytical formulas to describe them.
Section~\ref{sec:numerical} contains the main new results of this paper,
namely, a thorough numerical investigation of the tails of gluon density fluctuations in an onium.
The final section~\ref{sec:conclusion} presents our conclusions.
Technical details on the numerical calculations are gathered in the Appendices.

%%%%%%%%%%%%%%%%%%%%%%%%%%%%%%%%%%%%%%%%%%%%%%%%

\section{\label{sec:model}Multiplicity in hadron-nucleus collisions}

We recall the picture of particle production in p-A scattering
introduced in Refs.~\cite{Liou:2016mfr,Mueller:2016xti}. We first relate the final state
particle multiplicity to the gluon number density in the Fock state of the incoming
hadron at the time of its interaction with the nucleus, before explaining
how the event-by-event fluctuations of the gluon density can be thought of.

\subsection{Relation to the gluon number density in the hadron}

Let us consider most generally the scattering of a dilute hadron,
such as a proton or a quarkonium (which may be
either a model for a hadron, or
an actual state of a virtual photon), off a large nucleus, occurring at an energy
corresponding to the total relative rapidity~$Y$, assumed large compared to~1.
The gluons in the Fock state of the hadron at the time of the interaction\footnote{%
  We recall that the Fock state of a highly boosted hadron is essentially made
  of {\it gluons}. States containing extra quark-antiquark pairs are subdominant, and can be neglected
  in the so-called ``leading-logarithmic'' approximation. (See below for a definition of the latter.)
  }
that have a transverse momentum smaller than the saturation
scale of the nucleus
undergo scatterings, which may put them on-shell with high probability.
The ones that have
a transverse momentum larger hardly interact, and thus do not pick up the energy
that would be needed to produce them: Therefore, they must recombine with other partons before
they reach the final state. Hence, naively, the number of
hadrons measured in the final state at a given rapidity $y_0$ with respect
to the nucleus, in a given event, is proportional to the number of gluons with transverse
momentum smaller than the saturation momentum $Q_s(y_0)$ of the nucleus in the corresponding
Fock state of the hadron~\cite{Mueller:2016xti}.
It is tantamount to the gluon number density integrated up to this momentum which carry
a specific momentum fraction of the hadron in the initial state.
We shall denote it by $x{\cal G}(x,Q_s^2(y_0))$. 
The ordinary gluon density  $x{G}$ would be equal to the mean of $x{\cal G}$ when averaged
over the events.

\begin{figure}[ht]
  \begin{center}
    \includegraphics[width=0.65\textwidth]{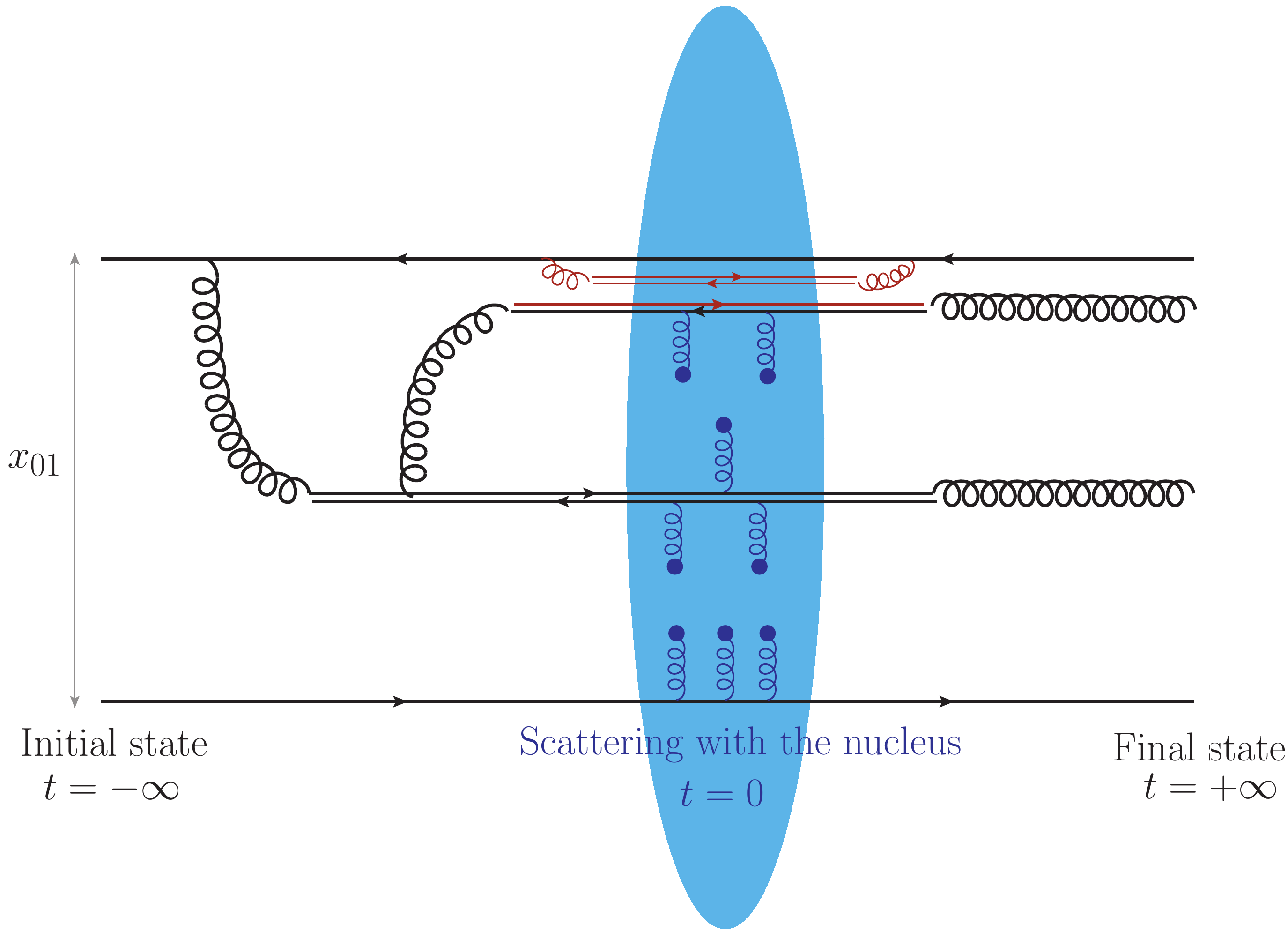}
  \end{center}
  \caption{\label{fig:scattering}\small
    {\it Space-time picture of a scattering event of an onium
    off a nucleus}. The frame is chosen such that the nucleus is leftmoving and
    boosted to the rapidity $y_0$, and the onium is rightmoving at
    rapidity $y\equiv Y-y_0$. In the particular event represented here,
    the onium fluctuates into 4 dipoles.
    Two of them are much larger than $1/Q_s(y_0)$ and interact
    with the nucleus, while the two others are much smaller than $1/Q_s(y_0)$ and do not interact.
    The scatterings are represented by gluon exchanges with the bulk of the nucleus.
    In a double-logarithmic scheme,
    only the gluons which are at the endpoints of the dipoles that scatter materialize in the
    final state, eventually in the form of hadrons of transverse momentum of the
    order of $Q_s(y_0)$. The other gluons recombine before they reach the final state.
  (The decay products of the nucleus are not represented).}
\end{figure}

More precisely, let us call
$M$ the mass of the onium and $dN/dy$ the number of gluons
per unit rapidity observed at an angle
corresponding to the rapidity $y_0$ relative to the nucleus (resp. $y\equiv Y-y_0$ relative
to the onium). Then, a calculation in the double-logarithmic approximation of
QCD leads to~\cite{Kovchegov:1998bi,Mueller:2016xti}\footnote{The relation was actually
  proven for the usual gluon density, namely averaged over events.
  We assume it holds
  true also for each event individually.
}
\be
\left.\frac{dN}{dy}\right.=x{\cal G}(x,Q_s^2(y_0))\quad\text{where}
\quad
x=e^{-y}\frac{Q_s(y_0)}{M}.
\label{eq:dNdy=xG}
\ee
A schematic representation of the mechanism behind the correspondence
formalized by this equation
is given in Fig.~\ref{fig:scattering}.
The gluons/hadrons produced in the final state have transverse momentum
of the order of $Q_s(y_0)$: Indeed, the multiple scatterings broaden the transverse momentum
of the gluons to that value.

Note that in the present model,
the saturation scale $Q_s(y_0)$ is a momentum which fully characterizes
the nucleus~\cite{McLerran:1993ni}
and depends only on the rapidity $y_0$, not on the considered event.
Indeed, since the nucleus
in its ground state is already a dense object,
the statistical fluctuations of its partonic content
can be neglected throughout its rapidity evolution.
The Fock state of the hadron instead results of a
stochastic evolution up to the rapidity $y$, starting with a few partons. Hence
it shows large event-by-event fluctuations,
which directly translate into large fluctuations of the multiplicity
observed in the final state.

Although our paper has a purely theoretical scope,
a comment on the phenomenological applicability of our results
and a justification of the relevance of our picture for proton-nucleus scattering at the
LHC is in order.
Our calculation requires a large nuclear saturation momentum $Q_s(y_0)$, of a few
GeV, in order for perturbation theory to be justified; Hence the 
rapidity $y_0$ should be relatively large.
The hadron carries the remaining available rapidity, $Y-y_0$, which
is assumed to be large, but small compared to rapidities at which
nonlinear effects should be taken into account in its evolution
(namely, parametrically, $Y-y_0\ll \ln^2 1/\alpha_s^2$,
see for example \cite{Enberg:2005cb,Iancu:2004iy,Munier:2014bba}).
This is consistent with the kinematics of LHC, where the beam rapidity is of order $Y\simeq 10$,
in such a way that we may pick
a suitable value of $y_0$, to allow a large-enough saturation momentum in the nucleus, and still
moderate evolution in the proton.

The simplest hadron we can think of is a quark-antiquark pair in a color
singlet state, which we call ``onium''. The Fock states of such an object are
most easily described, analytically and numerically, in the framework
of the color dipole model. Let us recall briefly how this works.

%%%%%%%%%%%%%%%%%%%%%%%%%%%%%%%%%%%%%%%%%%%%%%%%

\subsection{\label{sec:coll}Event-by-event fluctuations of the gluon number density}

When probed in its restframe with a wave of wavelength of the order of its spatial extension,
an onium is viewed as a bare quark-antiquark pair: Indeed, although they are ubiquitous,
the quantum fluctuations
are too short-lived on the scale of the interaction time to play a role in the interaction.
If instead the onium is probed with the same wave in a frame in which it has
a large rapidity, then the lifetimes of its quantum fluctuations
are Lorentz-dilated: Therefore, it appears essentially as a set of a large number of gluons.

One can evaluate the probability of a particular Fock state
at a given rapidity
by computing all diagrams
contributing to the probability amplitude of finding the onium in that state.
In the limit of a large number of colors $N_c$, and in the
leading logarithmic approximation (LLA) in which one keeps only the contributions
for which the number of powers of $y$ accompanying each power of $\alpha_s$ is maximum,
a convenient way to organize the calculation is the so-called color dipole model~\cite{Mueller:1993rr}.

The dipole model
uses coordinates in the two-dimensional plane orthogonal to
the worldline of the onium, instead of momenta, to
label the partons in the Fock state.
Thanks to the large-$N_c$ limit, the set formed by the initial
quark-antiquark pair along with its gluon fluctuations
can be replaced by a set of color dipoles, the endpoints of which coincide with
the position of the quark, of the antiquark, or of one of the gluons.
The graphs contributing to
a given state are generated by a stochastic branching process
in rapidity~\cite{Mueller:1993rr}. The latter is completely defined
by the elementary probability that a dipole defined by the pair of the
two-dimensional position vectors of its endpoints,
$(\uvec{x}_{0},\uvec{x}_1)$,
branches into two dipoles, $(\uvec{x}_0,\uvec{x}_2)$
and $(\uvec{x}_2,\uvec{x}_1)$ respectively, by emitting
a gluon at position $\uvec{x}_2$ when its rapidity is increased
by the infinitesimal amount $dy$. A calculation in the framework of perturbative
QCD leads to the following expression for the probability~\cite{Mueller:1993rr}
\be
dp_0|_\text{pQCD}=\bar\alpha dy
\frac{d^2\uvec{x}_2}{2\pi}\frac {{x}_{01}^2}{{x}_{02}^2 {x}_{12}^2},
\ee
where $\bar\alpha\equiv\alpha_s N_c/\pi$, and we introduce the notation
${x}_{ij}=|\uvec{x}_{ij}|$, with $\uvec{x}_{ij}=\uvec{x}_{i}-\uvec{x}_{j}$.
Analyzing this equation, one sees that there is a non-negligible probability that the
gluon be emitted at a large distance of the initial onium. However, 
confinement should forbid, at least in principle, the production of dipoles which are
bigger than, typically,~$1/\Lambda_\text{QCD}$. Being an intrinsically non-perturbative effect, we cannot
attain it through a perturbative calculation. Therefore, we add it to the original model
in the form of a ``cutoff function'' $\Theta$ that forbids dipoles of size typically larger than
some infrared length scale $\Rir\sim 1/\Lambda_\text{QCD}$ to be produced.
This leads to the following modification of the splitting probability:
\be
dp_0|_\text{pQCD}\longrightarrow
dp_0=\bar\alpha dy \frac{d^2\uvec{x}_2}{2\pi}\frac {{x}_{01}^2}{{x}_{02}^2 {x}_{12}^2}
\Theta(x_{02},x_{12};\Rir).
\label{eq:dp0}
\ee
Most generally, the cutoff function $\Theta$ must satisfy the following limits:
\be
\Theta(x_{02},x_{12};\Rir)\rightarrow
\begin{cases}
    0 & \text{if}\ x_{02}\gg \Rir\
\text{or}\ x_{12}\gg \Rir\\
1 &
\text{if}\ x_{02}\ll \Rir\ \text{and} \ x_{12}\ll \Rir.
\end{cases}
\ee
It is also expected to reach~0 ``fast enough'' (that is, at least exponentially) when
$x_{02}$ or $x_{12}$ become larger than $\Rir$.
This function is arbitrary in our treatment, however, it will become clear
that the observables we are interested in can depend only marginally on
its precise form.

The relation between the number of dipoles and
the gluon density is very simple
if one restricts oneself to double-logarithmic accuracy:
\be
x{\cal G}(x,Q_s^2)=\left.\frac{\partial}{\partial y}n(r_s=1/Q_s;x_{01},y)\right|_{y=\ln 1/x},
\label{eq:xG=n}
\ee
where $n(r_s;x_{01},y)$ is the number of dipoles of size larger than $r_s$ in the state
of an onium of initial size $x_{01}$, observed at rapidity~$y$.
The derivative enters the right-hand side because $x{\cal G}(x,Q_s^2)$ is the density
of gluons of a {\it fixed} momentum fraction $x$, while $n(r_s;x_{01},y)$ enumerates
the dipoles which have a rapidity {\it smaller} than~$y$.
It is the double-logarithmic approximation that enables one to identify the size
$r_s$ to the inverse momentum $1/Q_s$;
Sizes and momenta being conjugate to each other through Fourier
transform, this identification does of course not hold in general.

Thanks to Eq.~(\ref{eq:xG=n}), $x{\cal G}$, and thus, through Eq.~(\ref{eq:dNdy=xG}),
the number of particles produced in a given rapidity slice in the final state,
have the same fluctuations as the number of dipoles in the Fock state of the onium
at the time of the interaction. Therefore, the scope of the following sections will be
to study first analytically, and then numerically, 
the probability $P_n(r_s;x_{01},y)$
to have $n$ dipoles of size larger than $r_s$ in the Fock state of the onium
after evolution
of a dipole of initial size $x_{01}$ over the rapidity interval $y$.

%%%%%%%%

\section{\label{sec:asymptotics}
  Tails of the dipole number distribution}

In this section, we study
analytically the high and low-multiplicity tails of the dipole number
distribution $P_n(r_s;x_{01},y)$, developing physical
pictures which will prove useful
for the interpretation of the numerical data.

``High'' and ``low'' are intended with respect to the {\it expected} multiplicity.
In both cases, we will assume that the dipole numbers are much larger than unity.
In this limit~$n\gg 1$,
the probability $P_n$, which is defined as a function of the integer $n$, can be thought of
as a function of a continuous variable, and thus as a probability {\it density}.
The {\it probability} to observe a number $n$ of dipoles in the interval $[n_1,n_2]$
 then reads $\int_{n_1}^{n_2}dn\,P_n$.

\subsection{High-multiplicity tail}

\subsubsection{Heuristics}

\paragraph{No infrared cutoff.}

Let us recall that in perturbation theory, in
the absence of an infrared cutoff, the rapidity-evolution of the expected number of dipoles
larger than some size $r_s$, starting from an onium of size $x_{01}$,
is governed by the Balitsky-Fadin-Kuraev-Lipatov (BFKL)
equation~\cite{Kuraev:1977fs,Balitsky:1978ic}
\be
\frac{\partial}{\partial y}n^{(1)}(r_s;x_{01},y)
=\int \frac{dp_0|_\text{pQCD}}{dy}
  \left[
    n^{(1)}(r_s;x_{02},y)+n^{(1)}(r_s;x_{12},y)-n^{(1)}(r_s;x_{01},y)
    \right],
\label{eq:BFKL}
\ee
  where the integration goes over the whole transverse plane.
  Denoting by $\bar\alpha \chi(\gamma)$, where
  \be
  \chi(\gamma)=2\psi(1)-\psi(\gamma)-\psi(1-\gamma)\quad\text{with}\quad
  \psi(x)=\frac{d\ln\Gamma(x)}{dx},
  \ee
  the eigenvalue of the kernel of the BFKL equation associated
  to the eigenfunction $(x_{01}^2/r_s^2)^\gamma$, we can
  write the solution of equation~(\ref{eq:BFKL}) as a continuous superposition of 
  the eigenfunctions
  weighted by $e^{\bar\alpha y\chi(\gamma)}$.
  The initial condition corresponding to one single dipole of size
  $x_{01}$ reads $n^{(1)}(r_s;x_{01},y=0)=\theta(x_{01}-r_s)$.
  The solution to the BFKL equation then reads
\be
n^{(1)}(r_s;x_{01},y)=\int\frac{d\gamma}{2i\pi\gamma}
\left(
  \frac{x_{01}^2}{r_s^2}
  \right)^{\gamma}
  e^{\bar\alpha y\chi(\gamma)}.
  \label{eq:n(1)full}
\ee
The leading behavior of $n^{(1)}$ at large rapidities is
given by a saddle point:
\be
\begin{split}
&n^{(1)}|_\text{sp}(r_s;x_{01},y)\simeq \frac{1}{\gamma_s}\frac{1}{\sqrt{2\pi\bar\alpha y\chi''(\gamma_s)}}
\left(\frac{x_{01}^2}{r_s^2}\right)^{\gamma_s}
e^{\bar\alpha y \chi(\gamma_s)},\\
&\text{where}\ \gamma_s\ \text{solves}\quad \bar\alpha y\chi^\prime(\gamma_s)+\ln x_{01}^2/r_s^2=0.
\end{split}
\label{eq:n(1)}
\ee

\begin{figure}[h]
\begin{center}
    \includegraphics[width=0.9\textwidth]{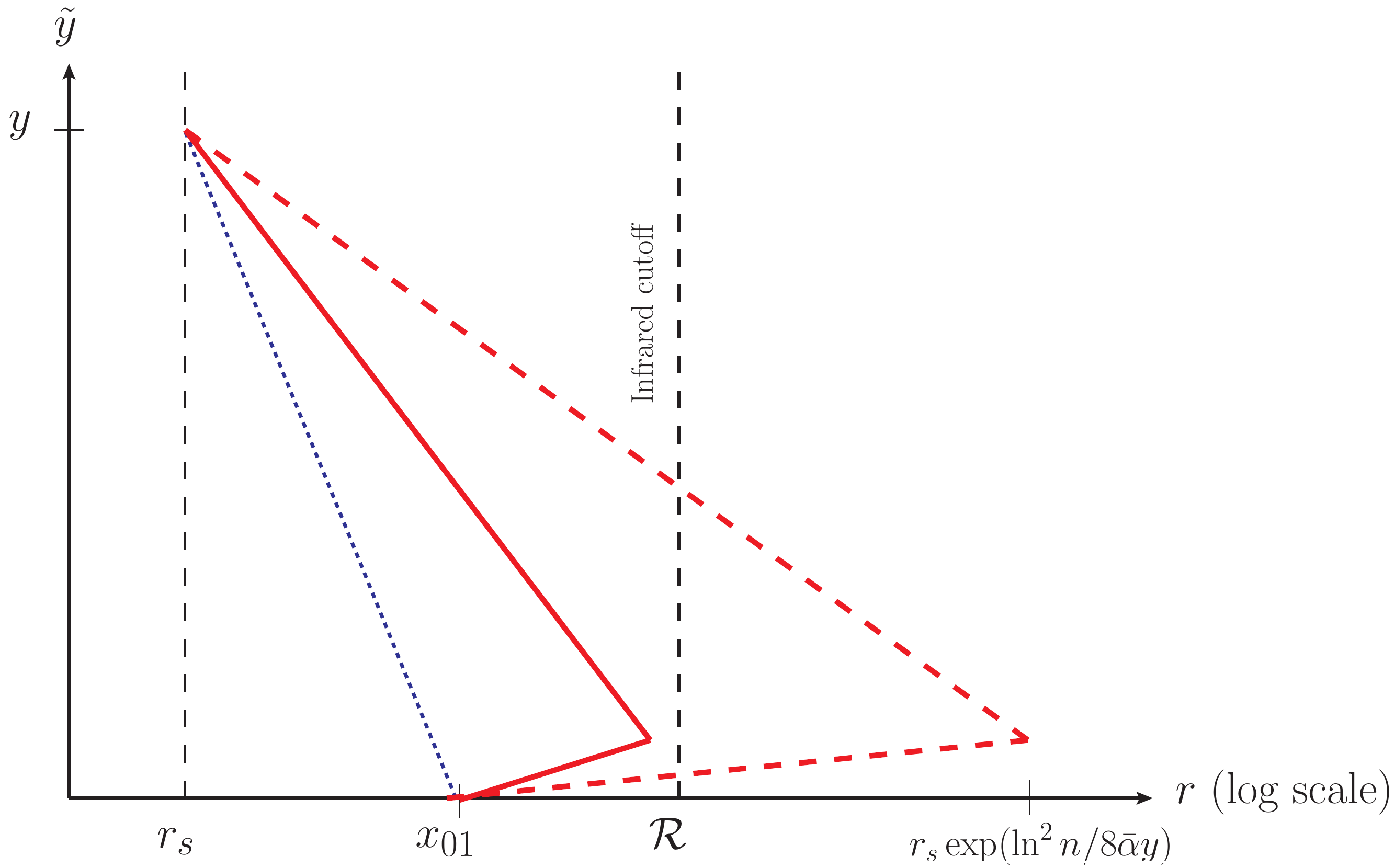}
  \end{center}
  \caption{\label{fig:evolution}\small
{\it  Evolution paths leading to typical or
    high-multiplicity dipole configurations.}
    The typical evolution can be represented as a straight line
    starting with the onium of size $x_{01}$ at rapidity $\tilde y=0$
    through the final size $r_s$ at rapidity $\tilde y=y$ (dotted line).
    In large-multiplicity rare events, there are two steps:
    in the first step, the initial onium rapidily evolves into a dipole of large size,
    function of the final multiplicity (dashed line), or of the order
    of the effective infrared cutoff ${\cal R}$
    in the presence of such a cutoff (continuous line).
    This large dipole subsequently
    decays into smaller ones through normal BFKL evolution (second step).
    }
\end{figure}

It is useful to represent the dipole evolution in the two-dimensional
$(r,\tilde y)$ plane as the curve
connecting the points $(x_{01},0)$ and $(r_s,y)$ and
that solves the saddle-point equation
\be
\tilde y=\frac{1}{\bar\alpha \chi'(\gamma_s)}{\ln r_s^2/r^2},
\ee
when the value of $\gamma_s$
is fixed by the boundary conditions, i.e. when it solves Eq.~(\ref{eq:n(1)}).
In log~$r$ scale, this curve is just a straight line,
see Fig.~\ref{fig:evolution}.
It represents the most probable evolution path.

If instead of the expected dipole number $n^{(1)}$
we are interested in the {\it probability} $P_n$
of having a given number $n$ of dipoles in the Fock state,
or if, equivalently, we focus on the set of moments $n^{(k)}$ of the dipole number, then the path
that corresponds to the main contribution
is not necessarily a straight line.
It was shown in Ref.~\cite{Liou:2016mfr}
that when $n$ is large compared to its
expected value (or equivalently, $k$ is large compared to~1),
this path essentially consists in two steps:
The initial onium generates, through a fast evolution,
a dipole of large size~$r_\text{max}$,
which subsequently decays into many (mainly smaller) dipoles.
The presence of the large dipole at an early stage
of the evolution is necessary 
if one asks for a large multiplicity,
because large dipoles yield much more offspring
of size larger than~$r_s$ than smaller ones.
This first step has a low probability, which decreases as~$\rmax$ increases.
But on the other hand, the number of offspring increases with~$\rmax$.
The optimal size $r_\text{max}$ of the intermediate large dipole
depends on the maximum rapidity and of the final number of dipoles of which
the probability is evaluated.

Following Salam in Ref.~\cite{Salam:1996nb},
we assume that the production of the large dipole occurs literally in the
very first step of the evolution. As we will check {\it a posteriori},
its size increases with~$n$,
and thus $r_\text{max}$ can be made arbitrarily large, say $r_\text{max}\gg r_s$,
by selecting very large values of $n$ at fixed $y$.
Once the large dipole has been produced, it decays into much smaller ones.
This second step in the evolution is dominated by
decays which are strongly ordered in the dipole sizes, from large to small.
The solution to the saddle-point equation in Eq.~(\ref{eq:n(1)}) is close to $\gamma_s\simeq 0$, a
region in which $\chi(\gamma_s)$ may be approximated by $1/\gamma_s$.
In this limit, the number of dipoles larger than $r_s$ resulting
from the decay of a dipole of size $\rmax\gg r_s$ reads
\be
n^{(1)}|_\text{DL}(r_s;\rmax, y)
\simeq e^{2\sqrt{\bar\alpha  y \ln \rmax^2/r_s^2}}.
\label{eq:DL}
\ee
This is the well-known ``double-logarithmic'' limit.
In Ref.~\cite{Liou:2016mfr}, it was proven that the fluctuations of the dipole number~$n$
on such an evolution path are suppressed exponentially.
We will check {\it a posteriori} that these fluctuations are overall negligible,
and thus, that we can assume
that the second step of the evolution is {\it deterministic}.
Consequently, the probability to observe more than say $N$ dipoles,
$\int_N^\infty dn\,P_n$, coincides with the probability that $\rmax$ be
larger than $R$, where $R$ is such that $n^{(1)}|_\text{DL}(r_s;R, y)=N$.
Solving this elementary equation for $R$, 
the relation between the probability distribution of the dipole number $n$ to that of the size of the
intermediate large dipole takes the following form:
\be
\text{Proba}(n\geq N)=
\text{Proba}\left(\rmax\geq r_s e^{\ln^2 N/(8\bar\alpha  y)}\right).
\label{eq:nsupN}
\ee

The probability that the initial dipole of size $x_{01}$
splits into a dipole of size $r$ larger than some given~$R$, itself much
larger than $x_{01}$, is suppressed
by the ratio of the squared sizes, see Eq.~(\ref{eq:dp0}).
Indeed, the distribution of the sizes of dipoles produced in the
splitting of a dipole of size $x_{01}$, conditioned to the occurrence of a splitting
into similar or larger-size dipoles
reads $\frac{1}{{\cal N}\bar\alpha}\frac{dp_0|_\text{pQCD}}{dy}$,
where ${\cal N}$ is a normalization factor of order~1. Thus the probability of having
a dipole of size $\rmax$ larger than $R$ is just the following integral:
\be
\text{Proba}(\rmax\geq R)\underset{R\gg x_{01}}\simeq
\frac{1}{{\cal N}}\int_{R}^{+\infty} \frac{d^2 \uvec{x}_2}{2\pi}\frac{x_{01}^2}{x_{02}^2 x_{12}^2}
\simeq \frac{1}{2{\cal N}} \frac{x_{01}^2}{R^2}.
\label{eq:probasplitlarge}
\ee
Now, to arrive at the distribution of the dipole number $n$,
it is enough to replace $R$ by $r_s e^{\ln^2 N/(8\bar\alpha  y)}$ (see Eq.~(\ref{eq:nsupN}))
in the previous equation.
Note that this quantity is the typical size of the intermediate dipole, which confirms
the {\it a priori} assumption made above that it
grows with $n$.
Taking finally the derivative with respect to $N$ and evaluating it at $N=n$,
we obtain the density\footnote{%
  Only the exponential factor is under control in Eq.~(\ref{eq:PnNoCutoff}):
  the other factors are not systematic.
  It was tested successfully against numerical simulations of the dipole model already
  in Ref.~\cite{Salam:1996nb}.
}
$P_n$:
\be
P_n=-\left.\frac{d\,\text{Proba}(n\geq N)}{dN}\right|_{N=n}
=\frac{1}{2{\cal N}}\frac{x_{01}^2}{r_s^2} \frac{1}{2\bar\alpha y}
\frac{\ln n}{n}\exp
\left({-\frac{\ln^2 n}{4\bar\alpha  y}}\right).
\label{eq:PnNoCutoff}
\ee
We note that the large-$n$ tail of $P_n$ is much fatter than an exponential decay.
This is the {\it a posteriori} justification
for having neglected the stochasticity in the second step of the evolution,
consisting in the decay of the large dipole of size $\rmax$.

A comment is in order. Dipole evolution is often assimilated with
a branching random walk (BRW). This is correct when one looks, for example, at
an observable probing the number of dipoles overlapping with a given point
in the transverse plane. But in a BRW, the number of objects
after some given evolution is distributed exponentially, by contrast with Eq.~(\ref{eq:PnNoCutoff}).
The fat tail we find here is a feature of QCD which shows up when we count {\it all}
dipoles (larger than a given size, to talk of an infrared-safe quantity) independently
of their transverse position. Technically, it is related to the size-dependence of
the dipole splitting rate, while in a BRW such as e.g. the branching Brownian motion, particle
splitting and diffusion are completely uncorrelated.
We are going to see that we actually recover an
exponential distribution when we enforce an infrared cutoff
on the evolution.

\paragraph{Enforcing an infrared cutoff.}

We now consider the modified dipole model which incorporates an infrared cutoff in the
form of the $\Theta$ function, see Eq.~(\ref{eq:dp0}).

With an IR cutoff, the typical size of the dipoles generated
in the first step of the evolution eventually becomes limited
by the infrared boundary if one focuses on very large values of $n$. Hence the
size~$\rmax$ of the intermediate dipole will always end up being of order $\Rir$.
So unlike in the purely perturbative QCD case, the stochasticity in $n$ cannot come from
the first step. It necessarily stems from the second step, consisting in the decay of
the large dipole.
Let us try to understand the form of the distribution of these fluctuations.

The IR cutoff forces the produced dipoles to be smaller than, typically, $\Rir$ throughout
the evolution. On the other hand, because it is probabilistically disfavored, a small dipole
does not split to much larger dipoles. So starting from a dipole of size close to $\Rir$,
the final number of dipoles is essentially built
up by a backbone of successive splittings of dipoles to similar-size or smaller,
but not {\it much} smaller, dipoles, each of which
gets dressed 
by a number of very small dipoles (of size of order $r_s$) proportional to the
rapidity interval over which it evolves.
Hence this second step in the evolution essentially looks like a $1\rightarrow 2$ branching process,
in which the branchings
occur at an almost constant rate, as in a BRW (see the comment above).
The fluctuations in such a process are known to be exponential,
$\propto e^{-n/n_1}/n_1$, where $n_1$ is the
mean number of objects eventually produced at the final rapidity.\footnote{
  In a $1\rightarrow 2$ branching process in time at fixed rate $r$,
  a straightforward calculation
  shows that the probability to have $n$ particles in the system at time
  $t$ reads $e^{-rt}(1-e^{-rt})^{n-1}$, when one starts with one particle at $t=0$.
  Since $\langle n\rangle=e^{rt}$, this probability also writes
  $e^{-n/\langle n\rangle}/\langle n\rangle$ when $n\gg 1$
  (see e.g. Ref.~\cite{Munier:2009pc}, Sec.~3).
  }
Hence, we expect the shape of the dipole number distribution to follow such a law.
We now need to understand the overall normalization, as well as the parameter~$n_1$.

Concerning the slope of the exponential, $n_1$, a good estimate can be obtained from the mean number
of dipoles produced by an initial dipole of size $\rmax$ of order $\Rir$.
It satisfies a modified
BFKL equation, i.e. Eq.~(\ref{eq:BFKL}) with the substitution
$dp_0|_\text{pQCD}\rightarrow dp_0$, which however cannot be solved exactly because
the eigenfunctions of the kernel of such an equation are not simple in general.
However, we may obtain a good approximation to its solution by replacing the $\Theta$ function
by a sharp Heaviside distribution, which in turn is tantamount to an absorptive boundary.
Then, the method of images can be used to arrive at a solution to this problem.

We start with the solution to the ordinary BFKL equation without a cutoff, Eq.~(\ref{eq:n(1)full}). We
may evaluate the integral in the saddle point approximation for large~$y$, using Eq.~(\ref{eq:n(1)})
with the substitution $x_{01}\rightarrow r$. 
We anticipate that the saddle point equation~(\ref{eq:n(1)})
for $\gamma_s$ has a solution near $\gamma_s=\frac12$,
therefore we replace
$\chi(\gamma)$ by its expansion around $\gamma=\frac12$:
\be
\chi\left(\gamma_s\right)\simeq
\chi\left({\scriptstyle \frac12}\right)+\chi'({\scriptstyle\frac12})
\left(\gamma_s-{\scriptstyle\frac12}\right)
+\frac12{\chi''({\scriptstyle\frac12})}\left(\gamma_s-{\scriptstyle \frac12}\right)^2=
4\ln 2+14\zeta(3)\left(\gamma_s-{\scriptstyle \frac12}\right)^2,
\ee
where $\zeta$ is the Riemann zeta function.
Then
\be
n^{(1)}|_\text{sp}(r_s;\rmax,y)\simeq
e^{\bar\alpha y 4\ln 2} \frac{\rmax}{r_s}
\left\{\sqrt{\frac{1}{\pi\bar\alpha y 14\zeta(3)}}\exp
\left[
  -\frac{\ln^2 (\rmax^2/r_s^2)}{\bar\alpha y 56 \zeta(3)}
\right]\right\}.
\ee
Following Ref.~\cite{Mueller:2002zm},
the absorptive boundary is implemented through
the method of images applied to the diffusive part, represented (up to numerical constants)
by the factor in the curly brackets in the previous equation.
The result reads
\begin{multline}
n^{(1)}|_\text{sp,$\Theta$}(r_s;\rmax,y)\simeq
e^{\bar\alpha y 4\ln 2}
\frac{\rmax}{r_s}\sqrt{\frac{1}{\pi\bar\alpha y 14\zeta(3)}}\\
\times\left\{
\exp
\left[
  -\frac{\ln^2 (\rmax^2/r_s^2)}{\bar\alpha y 56 \zeta(3)}
  \right]
-\exp
\left[
  -\frac{\ln^2 ({\Rir}^4/\rmax^2r_s^2)}{\bar\alpha y 56 \zeta(3)}
  \right]
\right\}.
\label{eq:n(1)boundary}
\end{multline}
It is easy to check that this solution obeys the (ordinary) BFKL equation, and
that it satisfies indeed the boundary condition $n^{(1)}|_\text{sp,$\Theta$}(r_s;\Rir,y)=0$.

We observe that there is an optimal dipole size, that maximizes the
mean number of dipoles at the end of the evolution:
$n^{(1)}|_\text{sp,$\Theta$}$ as a function of~$\rmax$ exhibits
a maximum at $\rmax={\cal O}(\Rir)$, located
between~$0$ and~$\Rir$. Indeed, it vanishes linearly with~$\rmax$
as $\rmax\rightarrow \Rir$ as a consequence of the presence of the absorptive
boundary, and also goes to zero as $\rmax\rightarrow 0$.

Concerning the normalization of the exponential,
the probability to generate a dipole of size~$\rmax$ of order ${\cal R}$
in the first step of the evolution
is a factor in $P_n$ independent of $n$ for large $n$.
It can be estimated by replacing $R$ by ${\cal R}$ in Eq.~(\ref{eq:probasplitlarge}):
The result is proportional to $x_{01}^2/{\cal R}^2$.

All in all, these heuristic considerations lead us to the following expression for $P_n$:
\be
P_n\propto \frac{x_{01}^2}{\Rir^2}
\frac{1}{n_1}e^{-n/n_1}.
\label{eq:heuristic}
\ee

This formula is indeed of the same form as the one derived
in Ref.~\cite{Liou:2016mfr} with a different, more mathematical, method.
The advantage of the present heuristic approach is that it comes
with a simple picture of the evolution of the Fock states into
high-multiplicities, while the more abstract method uses
the factorial moments, for which it may be more difficult
to build an intuition.
Also, we see that we have not used any detailed property
of the cutoff function $\Theta$: It just needs to be ``sharp enough'',
namely it must decay faster than some power of ${\cal R}/x_{02}$,
when $x_{02}$ (and thus $x_{12}$) gets larger than $\Rir$.
This shows that the high-multiplicity tail of the distribution of $n$
cannot be very sensitive to the precise form of $\Theta$.

Note that there is actually an awkward point in our heuristic discussion.
Indeed,  when the initial dipole has a small size $x_{01}$ compared to the infrared cutoff~$\Rir$,
if the production of the large-size dipole really consisted in
one single splitting, then
due to the geometry of dipole splitting,
one would have {\it two} dipoles of sizes $x_{02}$ and $x_{12}$
very close to each other, and not one.\footnote{A similar problem arose
  in the phenomenological studies of front fluctuations of Refs.~\cite{Brunet:2005bz,Mueller:2014gpa}.
  The so-called ``tip fluctuations'' studied in there, which are similar to the fluctuations to large
  dipoles in the present work, have always been assumed to consist in a single object produced
  in one step, independently
  of the model, although such an assumption is in general difficult to
  justify. But this effective description proved
  to lead to accurate analytical results in the context of
  Refs.~\cite{Brunet:2005bz,Mueller:2014gpa}, for reasons
  that have not been clarified so far.
}
Indeed,
$\uvec{x}_{01}=\uvec{x}_{02}+\uvec{x}_{21}$, so if $x_{02}\sim{\Rir}\gg x_{01}$,
then $x_{12}\sim x_{02}$.
But in a situation in which $x_{02}=x_{12}$, a simple calculation shows that
the fluctuations of the number of offspring of this {\it pair} of dipoles
would be distributed as $\propto n\times e^{-n/n_1}$ instead of a simple exponential.
As we will see in the detailed simulation of Sec.~\ref{sec:numerical},
this would contradict our numerical results.
There may be two ways out. First, the two dipoles are never exactly of the same size,
and consequently, the mean dipole yields
$n_1$ associated to each of them
are not exactly the same (see Eq.~(\ref{eq:n(1)boundary})).
Then, for very large $n$, the fluctuations are
always dominated by the offspring of one of the dipoles
(the one that yields most offspring on the average).
Second, in the more detailed analysis
of Ref.~\cite{Liou:2016mfr}, the production of the large dipole
resulted from a BFKL-like evolution, not
from one single splitting (although that evolution turned out to be very fast
when $n$ was set to be very large). In this case, there is no reason why there should
systematically be two large dipoles of (almost) identical size.
Finally, the exponential decay (without a $n$-dependent prefactor)
of $P_n$ with $n$ was found in the more straightforward calculation
presented in Ref.~\cite{Liou:2016mfr}
(and reproduced, for completeness, in the next section), and it
seems well supported by the numerical data, see below Sec.~\ref{sec:numerical}.

\subsubsection{Solution from an {\it\bfseries Ansatz}.}

We introduce the generating function $Z(r_s;x_{01},y|u)$
of the factorial moments of the dipole number:
\be
Z(r_s;x_{01}, y|u)=\sum_{n=1}^\infty u^n P_n(r_s;x_{01}, y).
\ee
It is well-known that it obeys the Balitsky-Kovchegov (BK)
equation~\cite{Balitsky:1995ub,Kovchegov:1999ua}
 (modified by the infrared cutoff here):
\be
\frac{\partial}{\partial  y}Z(r_s;x_{01}, y|u)
=\int \frac{d^2\uvec{x}_2}{2\pi}\frac{x_{01}^2}{x_{02}^2 x_{12}^2}\Theta(x_{02},x_{12};\Rir)
\left[Z(r_s;x_{02}, y|u)Z(r_s;x_{12}, y|u)-Z(r_s;x_{01}, y|u)\right].
\label{eq:BK}
\ee
For the analytic calculation, we choose a factorized form for $\Theta$:
\be
\Theta(x_{02},x_{12};\Rir)=
\tilde\theta\left({x_{02}}/{\Rir}\right)
\times
\tilde\theta\left({x_{12}}/{\Rir}\right),
\ee
where the function $\tilde\theta$ has the limits
\be
\tilde\theta(X)\xrightarrow{X\ll 1}1
\quad\text{and}\quad
\tilde\theta(X)\xrightarrow{X\gg 1}0.
\ee
Its precise form is not really relevant for the asymptotic calculations
we will carry out, except for one step (see below),
for which we will need to pick
a specific function for $\tilde\theta$, 
to arrive at a simple expression.

Even in the case of purely perturbative QCD (${\cal R}=+\infty$ or
equivalently~$\tilde\theta=1$),
we do not know how to solve the BK equation~(\ref{eq:BK}) accurately enough to be
able to extract the probabilities $P_n$ from the solution for~$Z$.
However, we may notice that the large-$n$ asymptotics of $P_n$ is connected to
the large-$k$ asymptotics of the factorial moments $n^{(k)}$.
It is straightfoward to convert the BK equation into a hierarchy
of equations for $n^{(k)}$:
\begin{multline}
\frac{\partial}{\partial  y}n^{(k)}(r_s;x_{01}, y)=
\bar\alpha\int\frac{d^2 {\uvec{x}_2}}{2\pi}
\frac{x_{01}^2}{x_{02}^2 x_{12}^2}\tilde\theta(x_{02}/\Rir)\tilde\theta(x_{12}/\Rir)
\bigg[
n^{(k)}(r_s;x_{02}, y)+n^{(k)}(r_s;x_{12}, y)\\
-n^{(k)}(r_s;x_{01}, y)
+\sum_{j=1}^{k-1}\left(\begin{matrix}{k}\\{j}\end{matrix}\right)
n^{(k-j)}(r_s;x_{02}, y)n^{(j)}(r_s;x_{12}, y)
\bigg],
\label{eq:n(k)cutoff}
\end{multline}
where the factorial moments $n^{(k)}$ are defined as the event-averages of the products
$n(n-1)\cdots(n-k+1)$, i.e.
\be
n^{(k)}=\left\langle \frac{n!}{(n-k)!}\right\rangle.
\ee
Again, it is not possible to find an explicit solution
for $n^{(k)}$ --- unsurprisingly, since the infinite set of equations~(\ref{eq:n(k)cutoff})
is equivalent to Eq.~(\ref{eq:BK}).
However, it is not difficult to figure out a plausible {\it Ansatz}.
We try
\be
n^{(k)}(r_s;x_{01}, y)=\frac{x_{01}^2}{\Rir^2}\tilde\theta(x_{01}/\Rir)C_k
\left[n^{(1)}(r_s;\Rir, y)\right]^k,
\label{eq:ansatz}
\ee
where the $C_k$'s are constants.

Taking $n^{(1)}=n^{(1)}|_\text{sp}$ from the saddle-point solution of the BFKL equation,
Eq.~(\ref{eq:n(1)}), inserting Eq.~(\ref{eq:ansatz}) into Eq.~(\ref{eq:n(k)cutoff}), and
keeping only the leading term in the limit of large rapidities, $n^{(1)}\propto e^{\bar\alpha y\chi(\gamma_s)}$,
the equation for the moments $n^{(k)}$ boils down to an equation for the constants $C_k$:
\begin{multline}
{\chi(\gamma_s)\tilde\theta(x_{01}/\Rir)} k C_k=
\bigg\{C_k\int\frac{d^2\uvec{x}_2}{2\pi}\frac{x_{01}^2}{x_{02}^2 x_{12}^2}
\tilde\theta(x_{02}/\Rir)\tilde\theta(x_{12}/\Rir)\\
\times\left[
\frac{x_{02}^2}{x_{01}^2}\tilde\theta(x_{02}/\Rir)
+\frac{x_{12}^2}{x_{01}^2}\tilde\theta(x_{12}/\Rir)
-\tilde\theta(x_{01}/\Rir)
\right]\bigg\}\\
+\sum_{j=1}^{k-1}
C_j C_{k-j}
\left(\begin{matrix}{k}\\{j}\end{matrix}\right)
\int \frac{d^2\uvec{x}_2}{2\pi \Rir^2}\tilde\theta^2(x_{02}/\Rir)\tilde\theta^2(x_{12}/\Rir).
\label{eq:nk+ansatz_0}
\end{multline}
The integrals over $\uvec{x}_2$ are finite functions of $x_{01}/\Rir$. We anticipate that the first terms in the right-hand side, inside the curly brackets, are negligible compared to the other terms. Under this assumption, which we will check a posteriori, the equation to solve simplifies to
\be
{\chi(\gamma_s)\tilde\theta(x_{01}/\Rir)} k C_k=
\sum_{j=1}^{k-1}
C_j C_{k-j}
\left(\begin{matrix}{k}\\{j}\end{matrix}\right)
\int \frac{d^2\uvec{x}_2}{2\pi \Rir^2}\tilde\theta^2(x_{02}/\Rir)\tilde\theta^2(x_{12}/\Rir).
\label{eq:nk+ansatz}
\ee
To push further the analytical calculation,
we now need an explicit form for the infrared cutoff function $\tilde\theta$.
We choose a Gaussian:
\be
\tilde\theta\left(X\right)=e^{-X^2/2}.
\label{eq:Gaussian_cutoff}
\ee
After the appropriate replacements have been done, the integration over $\uvec{x}_2$
in Eq.~(\ref{eq:nk+ansatz}) can
be performed:
\be
\begin{split}
\int \frac{d^2\uvec{x}_2}{2\pi \Rir^2}e^{-(x_{02}^2+x_{12}^2)/\Rir^2}
&=e^{-x_{01}^2/\Rir^2}\int_0^{+\infty} \frac{dx_{02}}{\Rir^2}
x_{02}\, e^{-2x_{02}^2/\Rir^2}
\int_0^{2\pi}\frac{d\theta}{2\pi}e^{2x_{01}x_{02}\cos\theta/\Rir^2}\\
&=e^{-x_{01}^2/\Rir^2}\int_0^{+\infty} \frac{dx_{02}}{\Rir^2}x_{02}\, e^{-2x_{02}^2/\Rir^2}
I_0(2x_{01}x_{02}\cos\theta/\Rir^2)\\
&=\frac14 e^{-x_{01}^2/(2\Rir^2)},
\end{split}
\ee
which is just $\tilde\theta(x_{01}/\Rir)/4$. Therefore,
Eq.~(\ref{eq:nk+ansatz}) boils down to an algebraic recursion for the constants $C_k$ of the form
\be 
kC_k=\frac{1}{4\chi(\gamma_s)} \sum_{j=1}^{k-1}\left(\begin{matrix}{k}\\{j}\end{matrix}\right)
C_j C_{k-j}~.
\ee
Inspection of this equation straightforwardly shows
that for asymptotically large~$k$, $C_k$ behaves like $a^k k!$, where $a$ is a constant.

We are now in a position to go back to Eq.~(\ref{eq:nk+ansatz_0}) and check {\it a posteriori} that it was indeed justified to neglect the terms in the curly brackets. This simply stems from the fact that these terms are overall proportional to $C_k$, while the sum of all the other terms, the ones we have kept, gives a contribution proportional to $k\times C_k$, which is much larger for $k\gg 1$.

Putting everything together, we see that the factorial moments of the dipole number read
\be
\label{eq:nkfinal}
n^{(k)}=c\times 4\chi(\gamma_s)\frac{x_{01}^2}{\Rir^2}e^{-x_{01}^2/(2\Rir^2)}k! n_1^k,
\ee
where $n_1=a\times n^{(1)}$
and $c$ is a constant which we expect to be of order~1.

Now, from the knowledge of the moments we can obtain the probability density function $P_n$.
Since we deal with typical multiplicities much larger than~1,
factorial moments of order $k$ can be approximated by ordinary moments of the same order, i.e.,
\be
n^{(k)}\simeq \langle n^k\rangle\simeq\int_0^\infty dn\,n^k\,P_n
\quad\implies\quad
P_n=\int \frac{dk}{2i\pi}n^{-k-1}n^{(k)}
\ee
which, using Eq.~(\ref{eq:nkfinal}), leads to the final result:
\be
P_n=c\times 4\chi(\gamma_s)\frac{x_{01}^2}{\Rir^2}
e^{-x_{01}^2/(2\Rir^2)}
\frac{1}{n_1}e^{-n/n_1}.
\label{eq:high_mult}
\ee
This is fully consistent with Eq.~(\ref{eq:heuristic}) found in the heuristic approach.
The extra Gaussian factor in Eq.~(\ref{eq:high_mult})
is dependent on the form of the cutoff function $\Theta$.

%%%%%%%%%%%%%%%%%%%%%%%%%%%%%%%%%%%%%%

\subsection{Low-multiplicity tail \cite{MuellerPrivate}\label{sec:lowmultTail}}

\begin{figure}[h]
\begin{center}
    \includegraphics[width=0.7\textwidth]{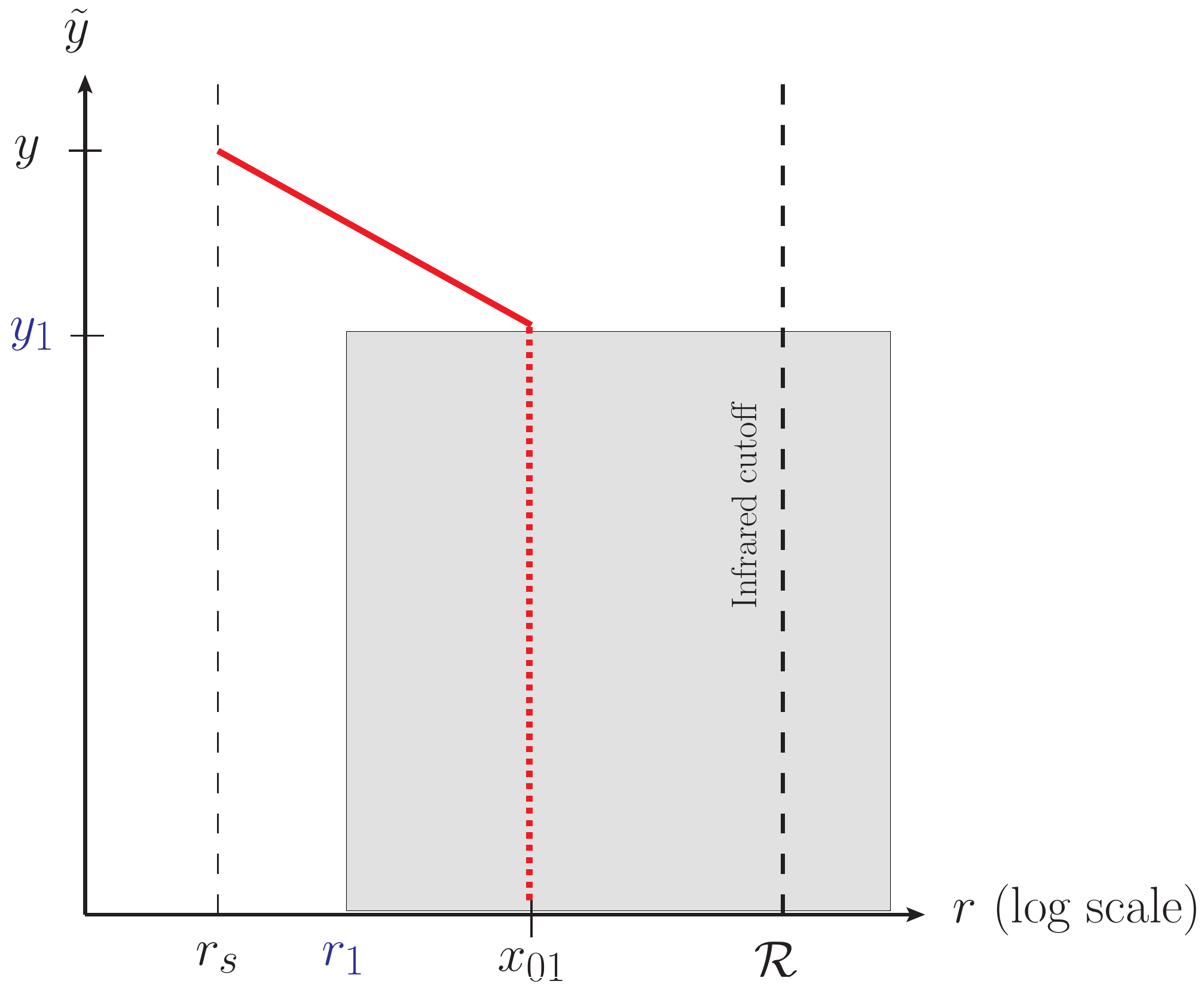}
\end{center}
\caption{\label{fig:evolution_low}\small
    {\it Evolution path leading to unusually low-dipole-number configurations.}
    The onium is prevented to split into large dipoles until a relatively large rapidity
    is reached.
    The forbidden region is represented by the shaded area. In that region, the evolution
    is limited to the emissions of small dipoles, of size smaller than~$r_1$.
    }
\end{figure}

We now turn to the evaluation of the distribution
of the low multiplicities. By ``low'' we mean much lower than the typical or mean
multiplicity~$\bar n\simeq n^{(1)}$, but at the same time still much larger than~1.
This region has not drawn as much attention as the high-multiplicity region,
with the exception of Ref.~\cite{Iancu:2003zr}, in which
Iancu and Mueller analyzed it
with a view to understanding the Levin-Tuchin
law~\cite{Levin:1999mw} for total dipole-nucleus versus dipole-dipole cross section
deep in the saturation region.

The only way to generate events with $n\ll \bar n$
is to veto the splittings of the initial dipole in the beginning of the evolution,
except if the latter are small enough:
Indeed, we know that
dipoles which have sizes close to the saturation radius~$r_s$ cannot evolve
into high-multiplicity states except
by creating large dipoles, but this has a large cost in probability
which makes such a process subdominant. On the other hand, once
the initial dipole has split into similar-size or larger dipoles,
then the cost of keeping the density of the state low becomes large.
Once a couple of dipoles have been emitted, the subsequent
evolution can be considered deterministic, and the latter generates a number of dipoles
which grows fast with the rapidity.
Hence we expect the low-multiplicity
tail of $P_n$ to be made of events in which the
occupation number is kept low throughout the initial stages of the evolution.
The evolution of such configurations is schematized in Fig.~\ref{fig:evolution_low}.

We are going to derive an expression for
the low-multiplicity asymptotics of $P_n$
from these simple considerations.
We shall assume that the probability of a given dipole number~$n$
coincides with a suppression factor for dipole splitting inside an appropriate region
${\cal D}$ of rapidity and transverse size, up to slowly-varying prefactors
that we shall discard:
\be
P_n\simeq\text{const}\times\exp\left(
-\bar\alpha\iint_{\cal D} d\tilde y\, \frac{d^2\uvec{x}_2}{2\pi}\frac{x_{01}^2}{x_{02}^2x_{12}^2}
\right).
\label{eq:Pn_low_0}
\ee
For $n$ small enough compared to the typical multiplicity $\bar n$,
${\cal D}$ must also include relatively small dipoles compared to $x_{01}$,
namely either $x_{02}\ll x_{01}$ or $x_{12}\ll x_{01}$. Let us call $r(\tilde y)$ the
lower boundary of ${\cal D}$ at some fixed $\tilde y$, namely the minium size of the dipoles
included in the domain over which we integrate.
Since the integral over ${\uvec{x}}_2$ diverges logarithmically when $r(\tilde y)$ goes
from ${\cal O}(x_{01})$ to zero, while the contribution of the dipoles larger than $x_{01}$ to
the integral is finite and of order~1,
keeping only
the strongly-ordered regions  $x_{02}\ll x_{01}$, $x_{12}\sim x_{01}$ and $x_{12}\ll x_{01}$, $x_{02}\sim x_{01}$
is enough to get the dominant term in the limit $r(\tilde y)\ll x_{01}$. We write 
\be
\int_{r(\tilde y)}\frac{d^2\uvec{x}_2}{2\pi}\frac{x_{01}^2}{x_{02}^2x_{12}^2}
\underset{r(\tilde y)\ll x_{01}}{\simeq} \int_{r(\tilde y)}^{x_{01}}\frac{dx_{02}}{x_{02}}+
\int_{r(\tilde y)}^{x_{01}}\frac{dx_{12}}{x_{12}}
=\ln \frac{x_{01}^2}{r^2(\tilde y)}.
\ee
Coming back to Eq.~(\ref{eq:Pn_low_0}), changing variable from $x_{02}$
to $\rho=\ln {x_{01}^2}/{x_{02}^2}$,
the following approximation
can be written for the probability:
\be
P_n\simeq \text{const}\times \exp\left(
-\bar\alpha\iint_{\cal D}d\tilde y d\rho\,\theta(\rho)
\right).
\ee

We pick the simplest for ${\cal D}$:
We assume it just consists in the rectangular-shaped region
$(r,\tilde y)\in[r_1,+\infty[\times [0,y_1]$.
We choose $y_1$ and $r_1$ such that a dipole of initial size $x_{01}$
starting to evolve deterministically at $ y_1$ produces exactly $n$
dipoles at the final rapidity $y$, and a dipole of size $r_1$ starting to evolve
at rapidity $0$ also produces $n$ dipoles at $y$.
It is not difficult to figure out that
these conditions are enough to guarantee that no dipole emitted outside of ${\cal D}$
can grow into a state of multiplicity much larger than $n$.
Hence in the presence of such a vetoed region, writing $\rho_1=\ln x_{01}^2/r_1^2$,
the distribution of the number of particles reads
\be
P_n\propto e^{-\bar\alpha  y_1 \rho_1}.
\label{eq:PnLow0}
\ee
Note that by choosing a rectangular region ${\cal D}$,
we neglect a term in $\ln P_n$ which is proportional to $(\bar\alpha y_1)^2$.

We now use the defining conditions for $y_1$ and $r_1$ to express these
variables with the help of
$n$ and $\bar\alpha y$.
In the double-logarithmic approximation~(\ref{eq:DL}),
these conditions read
\be
e^{2\sqrt{\bar\alpha(y-y_1)\rho_s}}=n
\quad\text{and}\quad
e^{2\sqrt{\bar\alpha y(\rho_s-\rho_1)}}=n,
\ee
where we introduced the notation $\rho_s=\ln x_{01}^2/r_s^2$.
The previous equations enable us to
rewrite Eq.~(\ref{eq:PnLow0}) as
\be
\ln P_n=\text{const}-\bar\alpha  y_1\rho_1=\text{const}-\bar\alpha y\rho_s
\left(
1-\frac{\ln^2 n}{4\bar\alpha y\rho_s}
\right)^2.
\ee
Finally, for the sake of writing down a more compact formula,
we may use again the double-logarithmic approximation
to express the product $\bar\alpha y \rho_s$ with the help of the expected
dipole number $\bar n$:
$2\sqrt{\bar\alpha y \rho_s}\simeq\ln \bar n$. The following
expression is obtained:
\be
\ln P_n=\text{const}-\frac14\ln^2\bar n\left(
1-\frac{\ln^2 n}{\ln^2\bar n}
\right)^2,
\ee
or, when expanded and ordered by decreasing importance in the limit $n\ll \bar n$:
\be
\ln P_n=\frac12\ln^2n-\frac14\frac{\ln^4 n}{\ln^2\bar n}
-\frac14\ln^2\bar n+\text{const}.
\label{eq:low_mult}
\ee

A remarkable feature of this result is
that the leading $n$-dependence at fixed $n$ and large rapidities
does not involve at all the infrared,
and actually does not depend on any scale at all.

%%%%%%%%

%%%%%%%%%%%%%%%%%%%%%%%%%%%%%%%%%%%%%%%%%%%%%%%%

\section{\label{sec:numerical}Numerical study}
In this section, we test the validity of the physical pictures proposed in Sec.~\ref{sec:asymptotics}
to understand the behavior of the tails of the multiplicity distribution.
To this aim, we perform high-statistics Monte Carlo simulations of dipole evolution
for different values of the parameters.
We measure distributions of the dipole multiplicity, $P_n$,
and compare their large and the small-multiplicity tails to Eqs.~(\ref{eq:high_mult})
and~(\ref{eq:low_mult}) respectively, for different values of the parameters.

Although we do not report on it here, we have also tested many choices of a rapidly falling $\Theta$,
and checked that the qualitative shape of the tails were not altered \cite{Domine:2017}, as expected
from general considerations.
In the numerical results we will present, we will restrict ourselves to the Gaussian IR cutoff
(Eq.~(\ref{eq:Gaussian_cutoff})) which was employed in the analytical calculations
of Sec.~\ref{sec:asymptotics}.

\subsection{High-multiplicity tail}

\begin{figure}[t!]
  \begin{center}
    \includegraphics[width=\textwidth]{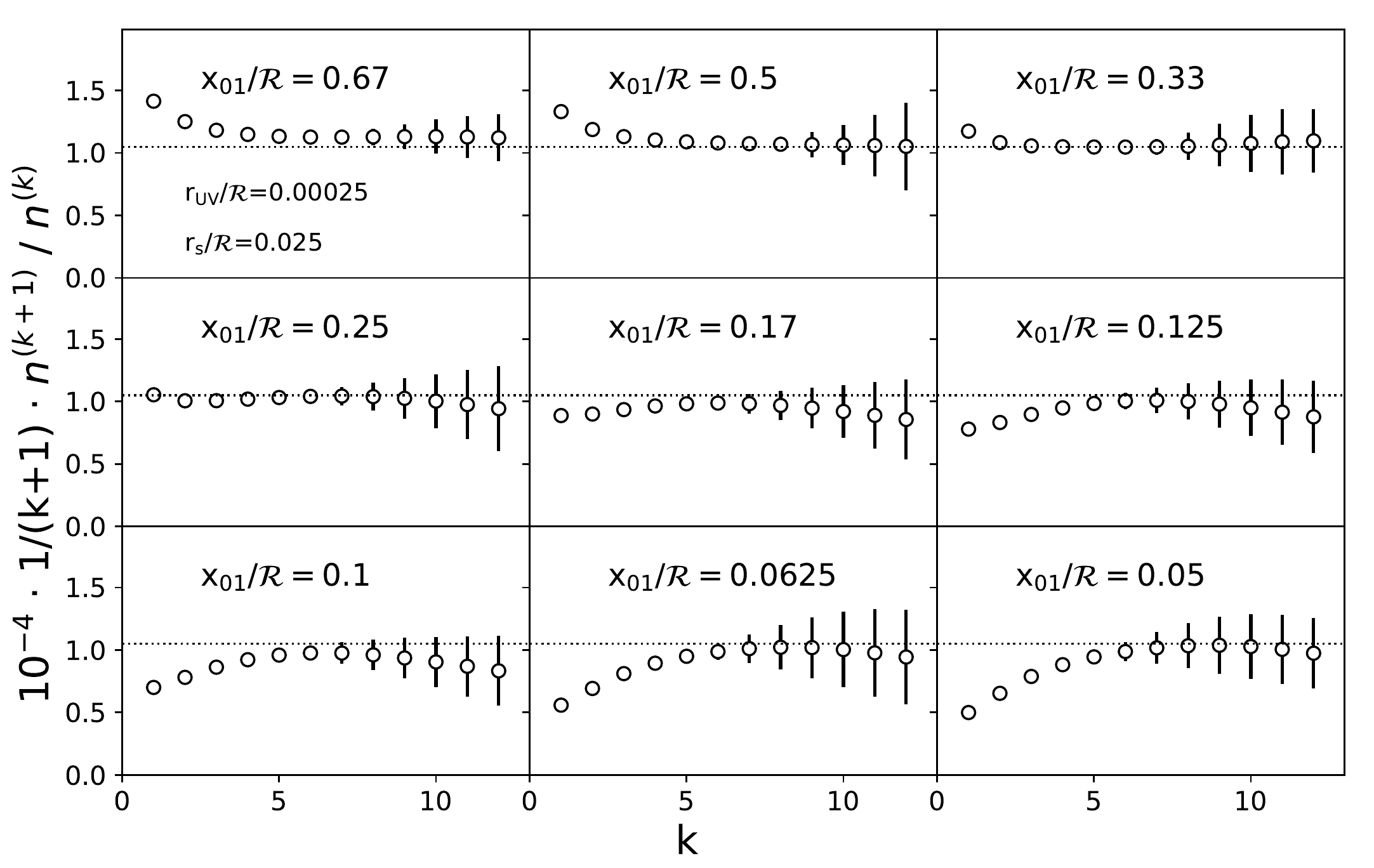}
  \end{center}
  \caption{\label{fig:n1}\small {\it Normalized ratio of moments of the dipole number distribution, $P_n$
      of successive order.} The ratio $n^{(k+1)}/[(k+1)n^{(k)}]$ is displayed
    as a function of $k$, for $\bar\alpha y=4$, and
    different values of $x_{01}/{\cal R}$. The dotted horizontal line is the constant $n_1$.
    The error bars are statistical.
  }
\end{figure}

The parameters of our numerical simulations are set to be the following:
$r_s/{\cal R}=1/40$, $x_{01}/{\cal R}$ are varied
between $0.67$ and $5\times 10^{-2}$, and $\bar\alpha y=2$ to $5$.
Our analytical results rely on the fact that the mean evolution between ${\cal R}$ and $r_s$
is driven by an eigenvalue of the BFKL equation close to $\chi(\frac12)$: In other words, the solution
to the saddle
point equation in Eq.~(\ref{eq:n(1)}) (with $x_{01}\rightarrow\Rir$)
was assumed to sit around $\gamma_s=\frac12$. Let us check that
it is indeed the case with the set of parameters we have chosen:
\be
\chi'(\gamma_s)=-\frac{\ln \Rir^2/r_s^2}{\bar\alpha y}\quad\implies\quad\gamma_s\simeq 0.45
\label{eq:eval_gs}
\ee
when $\bar\alpha y$ is set to be the value for which we have collected most of the data,
namely $\bar\alpha y=4$.

\subsubsection{Higher-order moments}

To study the behavior of the high-multiplicity tail, we look at the higher-order moments of~$P_n$.
In particular, for a given moment $n^{(k)}$, we look at its behavior as function of $k$, for different values of the size of the initial dipole, $x_{01}$.
Our goal is to check that $P_n$ has an exponential tail, and to provide a measurement of its slope.

If $P_n$ were a strict exponential distribution of the form, say, $P_n^{\text{exp}} = e^{-n/n_1}/n_1$,
then its moments would simply read
\be
n^{(k)}_\text{exp}=\int_0^\infty dn\,n^k\,\frac{1}{n_1}e^{-n/n_1}=k!\, n_1^k.
\label{eq:nexp}
\ee
the ratio of successive moments would satisfy the following equality:
\begin{equation}
\label{eq:ratio}
\frac{1}{k+1}\frac{n_\text{exp}^{(k+1)}}{n_\text{exp}^{(k)}} = n_1.
\end{equation}
Since the large-multiplicity
tail of the distribution is probed by moments of high order, we look at the behavior of this ratio
for large~$k$.

It is instructive to estimate the typical value of $n_s$ probed by a moment of order $k$.
This is given by the value of the dipole number,
$n$, that contributes most to the integral in Eq.~(\ref{eq:nexp}).
When $k$ is large, a saddle point at $n_s=kn_1$ dominates the integrand. 
These are the typical values of $n$ probed by the $k$-th moment.

Numerical results for Eq.~(\ref{eq:ratio}) in the case of initial dipoles of different relative sizes $x_{01}/{\cal R}$, evolved up to $\bar\alpha y=4$, are shown in Fig.~\ref{fig:n1}.
The very smooth behavior of the data points and of the magnitude of the statistical errors for different values of $k$ is due to an intrinsic correlation of errors at different $k$, which comes from the fact that for a given value of $x_{01}/\Rir$, $n^{(k)}$'s are computed using the same sample of data.
We draw a horizontal line at $n_1=10 500$ as an illustrative value of $n_1$ that is compatible, within one sigma, with all the curves shown in the plot
for large-enough~$k$.
We conclude that this value is independent
of $x_{01}/\Rir$.

\begin{figure}[t!]
  \begin{center}
    \includegraphics[width=\textwidth]{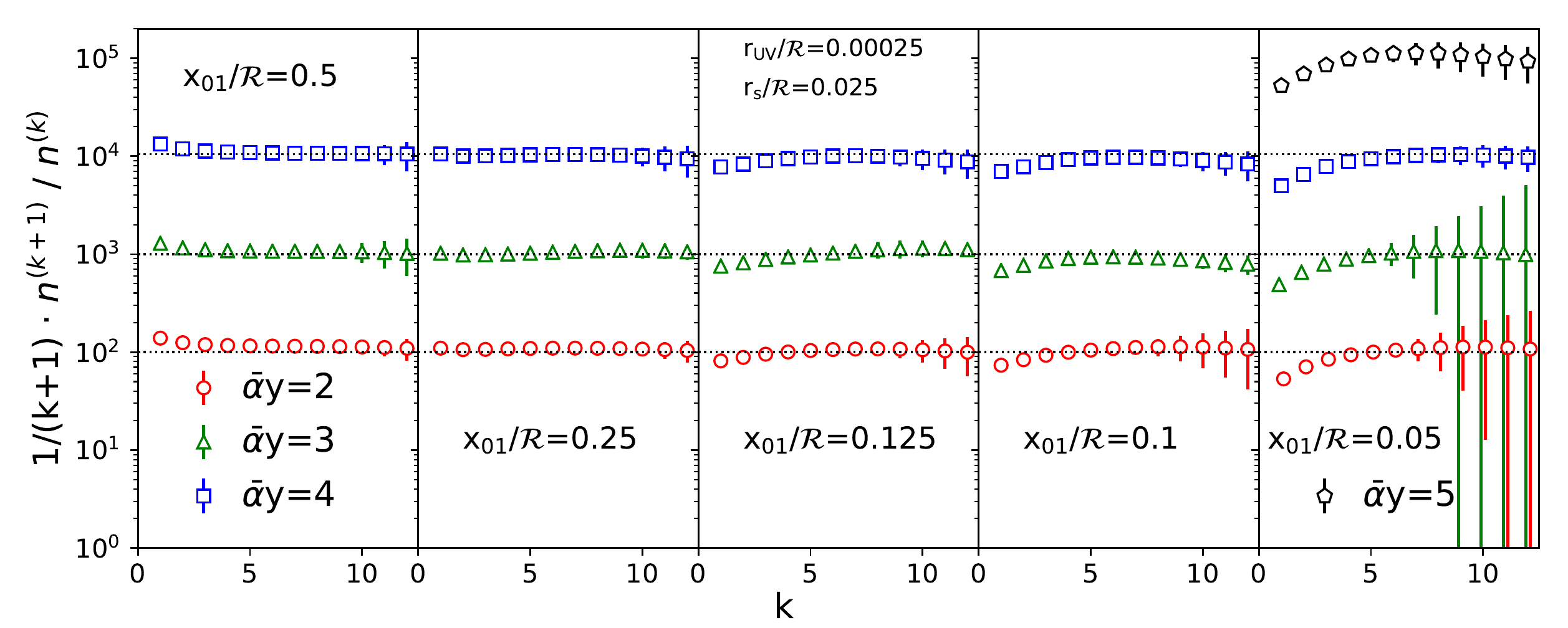}
  \end{center}
  \caption{\label{fig:n1_y}\small
    {\it Normalized ratio of the moments of $P_n$ of successive order.} The ratio $n^{(k+1)}/[(k+1)n^{(k)}]$ is dispayed
    as a function of $k$, for
    different values of $x_{01}/\Rir$ and different rapidities.
    The results for $\bar \alpha y=4$ are the same presented in Fig.~\ref{fig:n1}. For $\bar\alpha y= 5$,
    numerical results with $x_{01}/\Rir>0.05$ are prohibitively difficult to obtain. 
  }
\end{figure}

We now check that the rapidity dependence of $n_1$ is essentially exponential, up to prefactors, as predicted by Eq.~(\ref{eq:n(1)}).
For selected values of $x_{01}/\Rir$, we report results at different values of $\bar \alpha y$ in Fig.~\ref{fig:n1_y}.
Note that for any value of $y$, this ratio tends to a constant $n_1(y)$, which is independent of $x_{01}/{\cal R}$.
The logarithmic scale on the $y$-axis makes it obvious that $n_1$ grows with $y$ approximatively like an exponential, in agreement with the asymptotic identification $n_1\sim n^{(1)}$.

\begin{figure}[t!]
  \begin{center}
    \includegraphics[width=\textwidth]{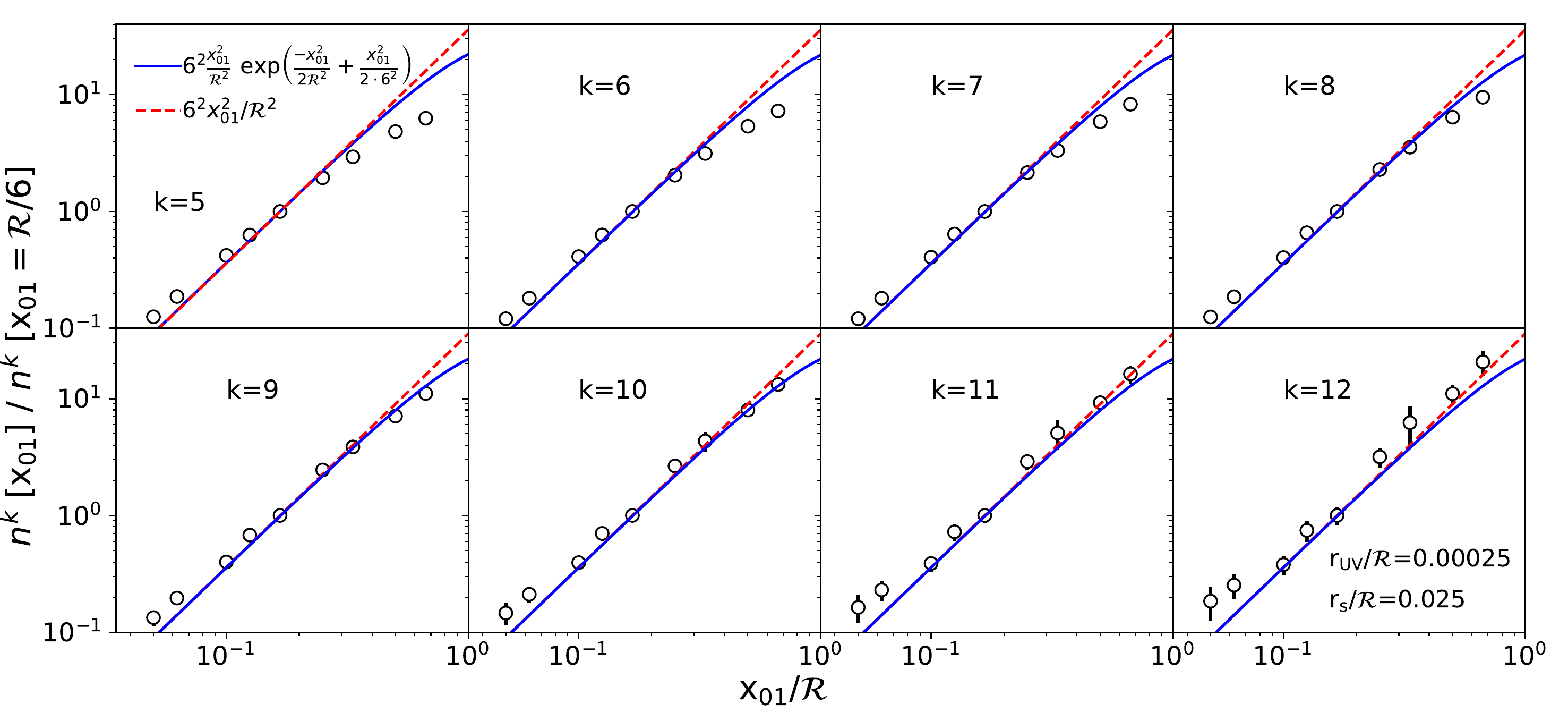}
  \end{center}
  \caption{\label{fig:coef}
    \small
        {\it Normalized moments of $P_n$ as a function of the onium size.} The $n^{(k)}$'s scaled by
        the constant $n^{(k)}[x_{01}=\Rir/6]$ are shown, for $k=5$ to $k=12$,
        as a function of $x_{01}/\Rir$, in logarithmic scale.
        The dashed straight lines correspond to the quadratic function $x_{01}^2\times 6^2/\Rir^2$
        and the curved full line to the same function corrected by a Gaussian factor, see Eq.~(\ref{eq:high_mult})
        and the legend.  Errors are statistical, and are computed via jackknife resampling.}
\end{figure}

Finally, our Ansatz predicts that the coefficient multiplying the exponential in Eq.~(\ref{eq:high_mult}) should present a specific quadratic dependence on the size of the initial dipole, $x_{01}$, for $x_{01} \ll \Rir$. 
In order to test this, we exploit the fact that this coefficient appears in the expression of the moments, Eq.~(\ref{eq:ansatz}). 
This implies that the ratio of two moments $n^{(k)}$ computed at two different values of $x_{01} /\Rir$ provides direct information about this coefficient. 
Hence, we compute the ratio $n^{(k)} [x_{01} / \Rir]/n^{(k)} [x_{01} / \Rir = 1/6]$ in our calculations at $\bar \alpha y = 4$. 
The results are shown in Fig.~\ref{fig:coef} up to $k=12$, after which statistical uncertainties dominate. 
We compare the numerical data with both a simple quadratic behavior (dashed line), and a quadratic Ansatz corrected with a Gaussian factor (solid line), i.e., the full prefactor in Eq.~(\ref{eq:high_mult}). 
We observe that the data points tend to fall on the expected curves\footnote{%
  In Fig.~\ref{fig:coef}, it seems that the two points corresponding to the lowest values of $x_{01}/\Rir$
  are systematically above the theoretical prediction, for all values of $k$. Actually this effect
  is not significant since, again, the values of $n^{(k)}$ for the different values of~$k$ have been
  calculated by averaging over the same sample of events, and thus, they are strongly correlated.
  } as we move to larger values of $k$.
Moreover, it is clear that the dashed and the solid line describe equally well the trend of the data points in the region where $x_{01}/\Rir$ is not close to unity.
We emphasize that this result is very important, because it implies that the simple heuristic discussion leading to Eq.~(\ref{eq:heuristic}) allows us to correctly predict the behavior of the numerical calculations.
This confirms the robustness of our intuitive picture of high-multiplicity events, and supports our statement that the exponential tail is universal irrespective of variations of the (sharp) infrared cutoff function.

In summary, all the expectations of Sec.~\ref{sec:asymptotics} are confirmed by our numerical results.
We conclude that, within our uncertainties, the high-multiplicity tail is an exponential correctly captured by Eq.~(\ref{eq:high_mult}).

\subsubsection{Shape of $P_n$ at large~$n$}

\begin{figure}[ht]
  \begin{center}
    \includegraphics[width=\textwidth]{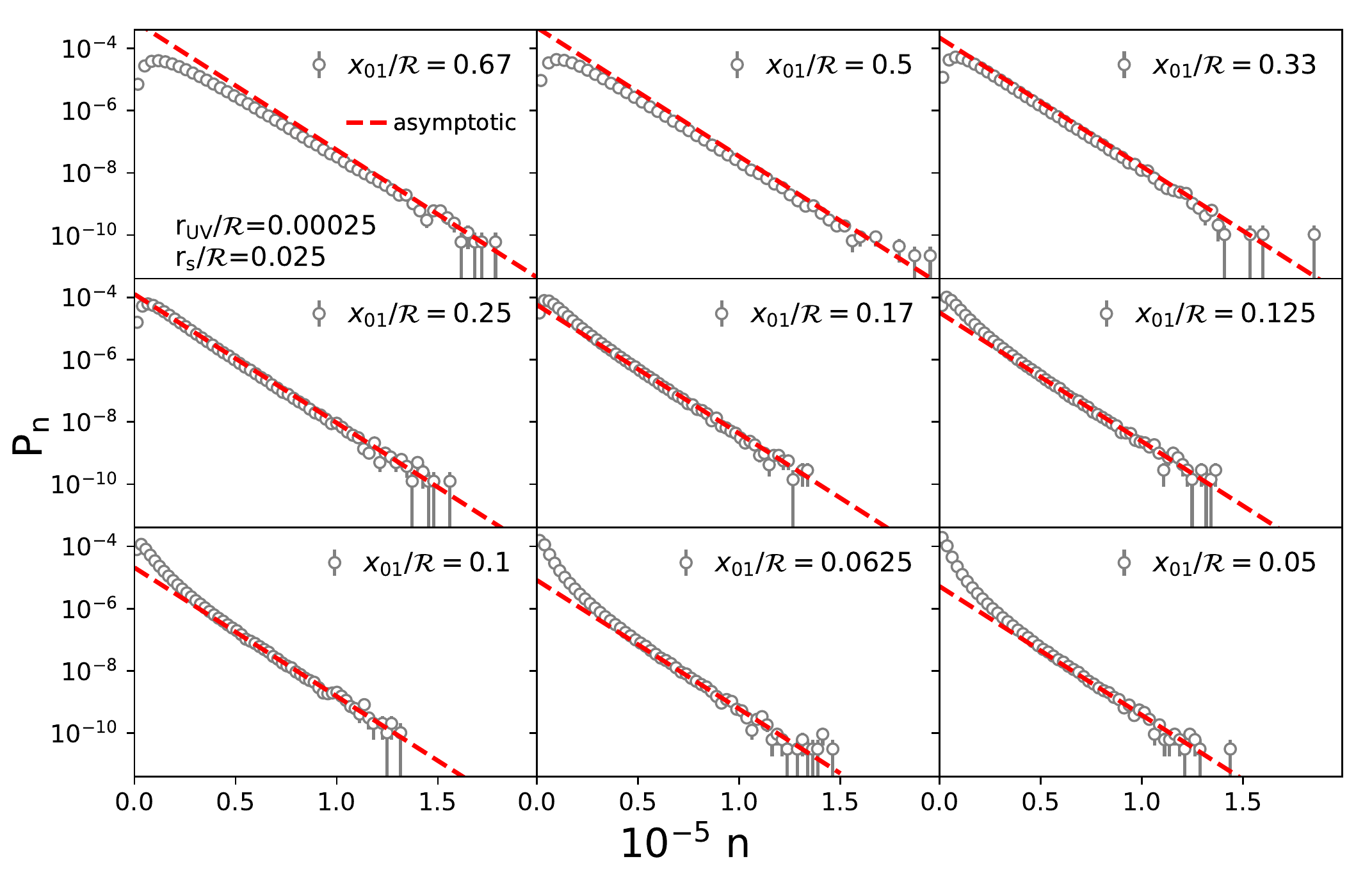}
  \end{center}
  \caption{\label{fig:P_n}\small {\it Distribution of the number of dipoles.}
    $P_n$ is displayed as a function of $n$ for
    different values of the onium size $x_{01}$, from $x_{01}/\Rir=2/3$
    to $x_{01}/\Rir=5\times 10^{-2}$,
    after evolution over $\bar\alpha y=4$.
    The straight dashed lines on this plot are the analytical asymptotics [Eq.~(\ref{eq:asympt})],
    with slope parameter $n_1=10500$, and global normalization $c=2$.
    In all cases, $\ruv/r_s=10^{-2}$, and $r_s/\Rir=1/40$.}
\end{figure}

Let us show, then, the actual shape of the distributions of dipole (gluon) number obtained in
our Monte Carlo calculations.
Results are shown as circles in Figure~\ref{fig:P_n}.
We exploit the result obtained in the previous subsection to characterize the tails at high multiplicity.
We overlay our curves with the asymptotics shown in Eq.~(\ref{eq:high_mult}),
\be
\label{eq:asympt}
P_n= c\times 4\chi(\gamma_s)~\frac{x_{01}^2}{\Rir^2} e^{-x_{01}^2/(2\Rir^2)}\frac{1}{n_1}e^{-n/n_1}
\tag{\ref{eq:high_mult}},
\ee
where we set $\gamma_s=\frac12$, and use $n_1=10500$,
i.e., the value inferred from the analysis of Fig.~\ref{fig:n1}, and we choose the overall normalization factor
to be $c=2$.
The asymptotic curves are shown as red dashed lines in Fig.~\ref{fig:P_n}.
We find that, with common values for $n_1$ and for the normalization, $c$, the exponential asymptotic trend is able to describe all the multiplicity distributions at large $n$, irrespective of $x_{01}/\mathcal{R}$.
We note that the constant~$c$ is indeed of order unity, as expected from the theory.

\subsubsection{Slope parameter $n_1$}

As already mentioned, the parameter $n_1$ in the exponential appearing in Eq.~(\ref{eq:asympt})
is expected to be related to the mean
number of dipoles larger than $r_s$ generated by the deterministic
evolution over $y$ units of rapidity of a dipole
of initial size $\rmax$ of the order of $\Rir$.
To check the consistency of this interpretation of~$n_1$,
let us compare the analytical expectations of such a scenario to the
numerical data for~$n_1$.
Putting numbers into Eq.~(\ref{eq:n(1)boundary}), we find
that for $\bar\alpha y=4$ and ${\Rir}/{r_s}=40$, $n_1$ 
reaches a maximum of about
11700 for $\rmax\simeq 0.37\times\Rir$.
This is consistent with the value of $n_1$ found in the numerical data
($n_1\sim 10500$).
Remarkably, it is for such values of $x_{01}$
that $P_n$ is closest
to a pure exponential, see Fig.~\ref{fig:P_n}.
This means that when one sets $x_{01}$ to the value $\rmax$ expected to maximize the number
of final dipoles,
then the optimal path in the $(r,\tilde y)$ plane (see Sec.~\ref{sec:asymptotics})
is a straight line.
The emission of large dipoles in the initial steps is no longer advantageous,
since the probability is strongly dumped by the cutoff function.
This is a nice consistency check of the whole picture.

We have checked that the $y$-evolution of $n_1$ which we may deduce from
Fig.~\ref{fig:n1_y}
is also qualitatively reproduced by Eq.~(\ref{eq:n(1)boundary}).

\subsection{Low-multiplicity tail}

\begin{figure}[t!]
  \begin{center}
    \includegraphics[width=\textwidth]{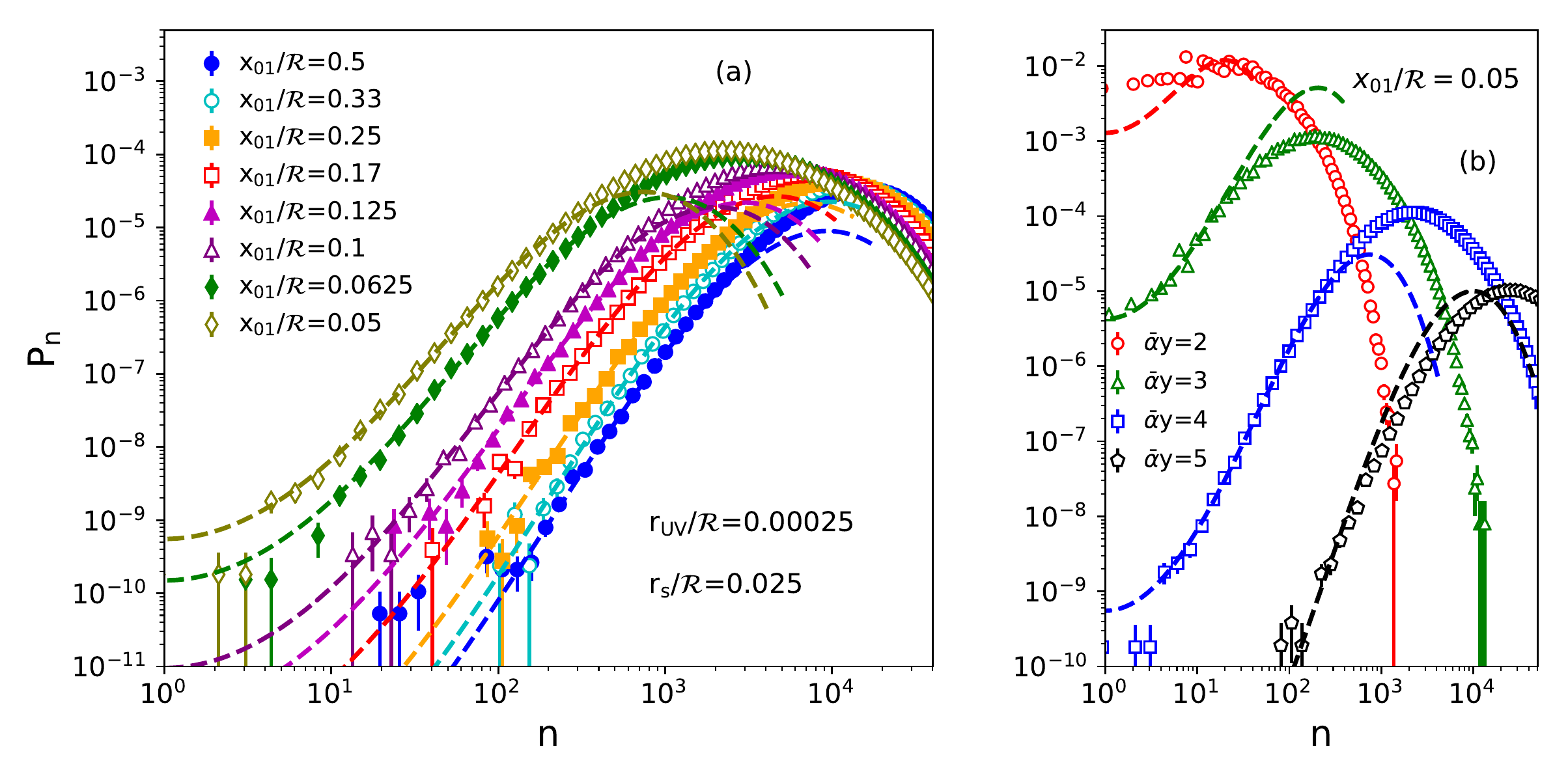}
  \end{center}
  \caption{\label{fig:low_mult}\small
    {\it Dipole number distribution on a logarithmic scale to emphasize the low-multiplicity tail.}
    {(a)} Distribution of the dipole number for different values of $x_{01}/\Rir$
    and fixed $\bar\alpha y=4$. Dashed lines are fits using Eq.~(\ref{eq:low_mult}).
    {(b)} The same but for different values of $\bar\alpha y$ and fixed $x_{01}/\Rir=0.05$.
  }
\end{figure}

Now we turn our attention to the low-multiplicity tail of the dipole number distribution, $P_n$.
Again, we want to check that the physical picture described in Sec.~\ref{sec:asymptotics} for the low-$n$ tail is consistent with the Monte Carlo results.
To this aim, we shall perform fits of $P_n$ in the low-multiplicity region using the formula in Eq.~(\ref{eq:low_mult}). 

Figure~\ref{fig:low_mult}(a) shows the results for the low-multiplicity tails of the distributions obtained at different values of $x_{01}/\mathcal{R}$, for $\bar \alpha y =4$.
The dashed lines in the figure represent fits obtained using Eq.~(\ref{eq:low_mult}).
The strategy of our fitting analysis is the following: For each tail, we fit the same number of points (around 15), starting from the points with lowest probability.
We checked that, doing so, we do not break the condition $n\ll\bar n$, which is required for Eq.~(\ref{eq:low_mult}) to apply.
We find that the quality of the fit improves as we move to larger values of $x_{01}/\mathcal{R}$, as the $\chi^2$ per degree of freedom (dof) of the fit is close to 4 at $x_{01}/\mathcal{R}=0.05$, and close to unity at $x_{01}/\mathcal{R}=0.5$.
This behavior confirms our expectation that the physical picture presented in Sec.~\ref{sec:asymptotics} works as long as $1\ll n\ll\bar n$, a requirement which is loosely satisfied by the curves at small $x_{01}/\mathcal{R}$, where we observe events with $n\sim1$.
For each tail, the two-parameter fits return one overall normalization, and a value for the mean of the distribution, $\bar n$.
Remarkably, we find that the overall normalization varies by less than a factor 2 among the different fits, as long as $x_{01}/\mathcal{R}>0.1$.
This supports our conclusion of a universal shape for the low-multiplicity tail, which is simply shifted towards larger values of~$n$ if $\bar n$ increases.
As for the fitted values of $\bar n$, we obtain numbers which are smaller than the actual mean value of the distribution, although in a systematic way: The fitted $\bar n$ at $x_{01}/\Rir=0.5$ is much closer to the true value than for the fit at $x_{01}/\Rir=0.05$.
Again, we understand this as a consequence of the fact that in our calculations we do never reach the fully asymptotic regime $1\ll n\ll \bar n$, since even for $x_{01}/\Rir=0.5$ we observe events with $n\sim10$.
These systematics support the validity of Eq.~(\ref{eq:low_mult}).

To corroborate this statement in more generality, it is useful to look at the results shown in Fig.~\ref{fig:low_mult}(b).
In this panel, we focus on the multiplicity distribution at $x_{01}/\Rir=0.05$, which is the distribution providing the least satisfactory results from the fit of the low-$n$ tail.
We show how the tail evolves with the rapidity, $\bar \alpha y$.
We see that increasing rapidity has an effect similar to that of increasing $x_{01}/\Rir$, in the sense that the distribution gets shifted towards larger values of~$n$.
%We simply obtain that the fit at $\bar \alpha y=5$ shown in Fig.~\ref{fig:low_mult}(b) is of the same quality as the fit for $\mathcal{R}/x_{01}$ shown in Fig.~\ref{fig:low_mult}(a),
%but we are more limited by statistics since results at $\bar \alpha y=5$ are very difficult to achieve.
Clearly, the quality of the fit improves as we increase the value of $\bar\alpha y$, except for the largest
$\bar\alpha y=5$, for which 
the trend of the data is less well reproduced by the theoretical curve. This is actually
expected, because
the double-logarithmic approximation, crucial in the derivation of Eq.~(\ref{eq:low_mult}), is less justified for larger values of $\bar\alpha y$.

\begin{figure}[t!]
  \begin{center}
    \includegraphics[width=0.7\textwidth]{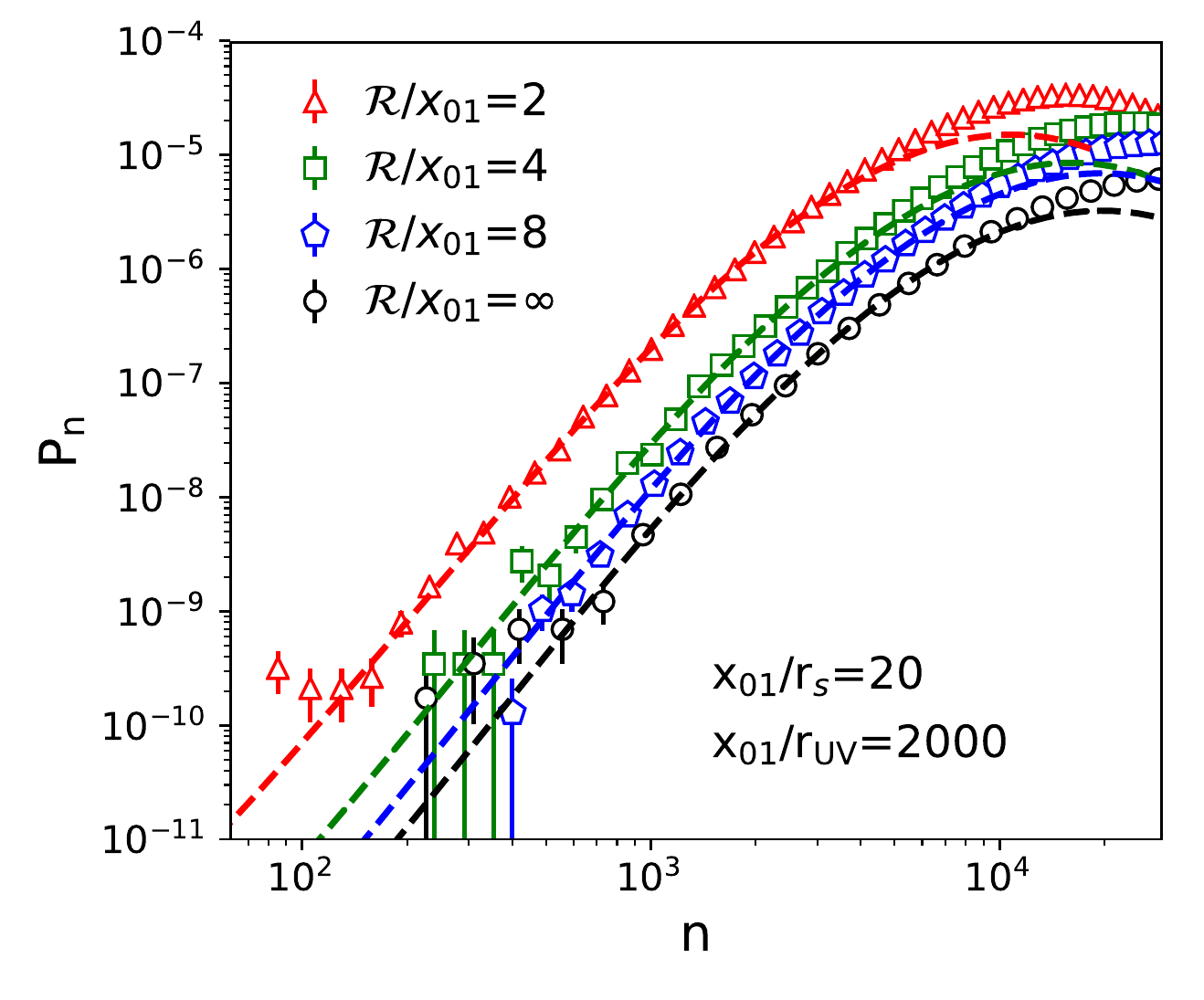}
  \end{center}
  \caption{\label{fig:low_mult_infty}\small
    The same as in Fig.~\ref{fig:low_mult}, but $x_{01}/r_s$ is now kept fixed, and different
    values of the infrared cutoff are taken. Again, $\bar \alpha y=4$.
  }
\end{figure}

We now show that the shape of $P_n$ in the low-multiplicity region is not affected by the presence of the infrared cutoff, $\Theta$.
To this aim, we perform calculations with fixed ratios $x_{01}/r_s$ and $x_{01}/r_{\rm UV}$, whereas we vary $\mathcal{R}/x_{01}$, i.e., the distance between the initial dipole size and the infrared cutoff.
We test very different scenarios: We consider a range of values for $\mathcal{R}/x_{01}$, as well as the case where the infrared cutoff is absent, i.e., $\mathcal{R}/x_{01}=+\infty$.
We obtain the results shown in Fig.~\ref{fig:low_mult_infty}.
The two-parameter fits, obtained from Eq.~(\ref{eq:low_mult}), are of excellent quality, and return a $\chi^2$/dof close to unity.
We conclude that the infrared cutoff has simply the effect of translating $P_n$ towards lower values of~$n$, without altering the shape of the low-multiplicity tail.

Note that the agreement we have found is better than one would have expected,
considering the values of the parameters we have chosen. Indeed, the double-logarithmic approximation was heavily used in deriving Eq.~(\ref{eq:low_mult}), and the latter is supposed to be best justified for $\gamma_s\rightarrow 0$, where $\gamma_s$ is evaluated using Eq.~(\ref{eq:eval_gs}),
with the substitution $\Rir\rightarrow x_{01}$. It turns out that $\gamma_s$ ranges between~0.41 and~0.46 when computed with the parameters with which our data has been generated.

\begin{figure}[t!]
  \begin{center}
    \includegraphics[width=\textwidth]{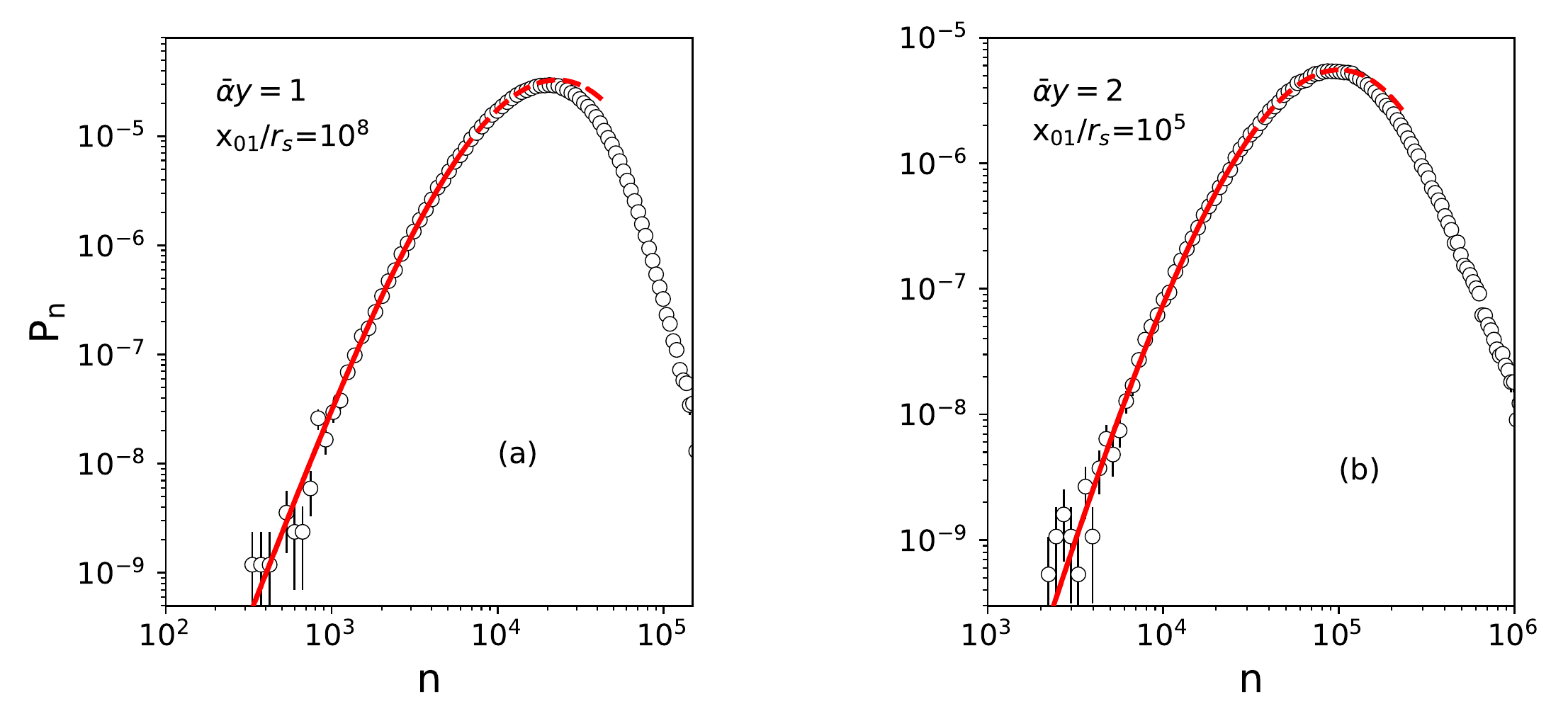}
\end{center}
\caption{\label{fig:low_mult2}\small
  {\it Dipole number density in log scale, for lower $\bar\alpha y$ and $r_s$.}
  The line, that represents the fit, is continuous in the range in which the data have
  been taken into account for the determination of the parameters, and dashed when it represents
  an extrapolation.
  In both plots, $\Rir=+\infty$.
    }
\end{figure}

Therefore, it is important to confirm that the analytical formula~(\ref{eq:low_mult}) works also
very well in the kinematical region which satisfies the assumptions
made for its derivation. To this aim, we generate data at lower rapidities and much larger
values of $x_{01}/r_s$:
\be
\begin{split}
  \bar\alpha y&=1\quad\text{and}\quad\frac{x_{01}}{r_s}=10^8\quad\implies\quad \gamma_s\simeq 0.16\\
   \bar\alpha y&=2\quad\text{and}\quad\frac{x_{01}}{r_s}=10^5\quad\implies\quad \gamma_s\simeq 0.28.
\end{split}
\ee
The corresponding values of $\gamma_s$ are now such that the DL approximation is
very well justified. The agreement between the numerical data and
the theoretical parametrization is
indeed extremely good.
Both fits return a value of $\chi^2/\text{dof}$ close to unity;
see Fig.~\ref{fig:low_mult2}.

In conclusion, the shape of gluon number distribution in the region of low multiplicity is not affected by the presence of an infrared cutoff, and it is correctly captured by the physical picture of Sec.~\ref{sec:lowmultTail}, as long as the parameters $\bar\alpha y$, $x_{01}$ and $r_s$ are such that there is a large-enough region in which the inequalities $1\ll n \ll \bar n$ are satisfied.

%%%%%%%%%%%%%%%%%%%%%%%%%%%%%%%%%%%%%%%%%%%%%%%%%%%%

\section{Summary and outlook\label{sec:conclusion}}

We have studied analytically and numerically the tails of the multiplicity distribution
of gluons corresponding to dipoles larger than a given size $r_s\sim 1/Q_s(y_0)$
in the Fock state of an initial onium.

The low-multiplicity tail shows very robust features, as its leading dependence in $n$ does
not depend on any scale or parameter. In particular, it does not depend on the infrared cutoff,
and we have been able to describe it with Eq.~(\ref{eq:low_mult}) even at the border of the double-logarithmic
region.
The large-multiplicity tail is exponential: A proxy for
the parameter of this exponential
is the mean integrated gluon density in a dilute hadron evaluated at the saturation
scale $Q_s(y_0)$.
The onset of the exponential depends on the ratio of the onium size to
the confinement size.

It is important to appreciate that the fluctuations of the gluon density,
which are relevant for the multiplicity distribution in hadron-nucleus
scattering we have studied, are fundamentally different
in nature from the event-by-event fluctuations of
the saturation scale discussed in Ref~\cite{Iancu:2004es,Hatta:2006hs}.
The source of these fluctuations was identified to be the saturation
of the dipole density in the evolution itself
(whose mechanism may be recombinations, or screening of further emissions), which
we have not implemented in the dipole model and which are not relevant
at the energies of the present colliders.
The fluctuations studied here are also different from
the so-called ``front'' and ``tip'' fluctuations in branching random walks identified in
Ref.~\cite{Mueller:2014fba,Mueller:2014gpa}.
The latter prove relevant in QCD for observables for which only dipoles overlaping with
a given impact parameter may play a role, or only the fluctuations of the size
of the largest dipole(s):
The total and diffractive $\gamma^*$-A deep-inelastic scattering cross sections are
examples of such observables, whose evolution can be considered
in the universality class of branching random walks.
The fluctuations which are at work in the process we have been analyzing in this paper
are essentially equivalent to the statistical noise of the total number of particles
generated by a $1\rightarrow 2$ branching process after some given evolution.

Going from our study of the multiplicity of gluons produced in onium-nucleus collisions to the
experimental data on the hadron multiplicities in proton-nucleus scattering
will require some modeling. First, the
details of the state of the
initial proton will certainly determine
the overall normalization, but may also introduce $n$-dependent prefactors. Second, a hadronization model
will be needed to link our parton-level calculation to the actual hadronic final state.
The modelization of the proton is presumably not easy.
Several recent studies model its color field as a classical Gaussian field,
see Ref.~\cite{Kovner:2018azs} and references therein.
We deem that a set of a few partons is more appropriate,
but whether a diquark dipole would be a
good representation for its groundstate is questionable.
Exclusive diffractive processes may help constraining the initial condition
for the evolution, see e.g. Ref.~\cite{Mantysaari:2016jaz}.

But while the detailed form of the multiplicity
distribution will certainly depend on the
model for the initial hadron, we expect
its general features to be robust.
An interesting observable that could be amenable to dipole model description
is multiplicity
in the final state of deep-inelastic $\gamma^*$-A scattering,
for sets of events in which the photon virtuality is chosen smaller than the nuclear
saturation scale. In such a case, the virtual photon interacts through its quark-antiquark
pair fluctuations, the distribution of the size of which is obtained from a straightforward
QED calculation.
Therefore, this process is very close to onium-nucleus scattering studied
in the present paper.

{
\section*{Acknowledgements}

The work of CL and SM is supported in part by the Agence Nationale
de la Recherche under the project \# ANR-16-CE31-0019.
We thank Prof. A.~H. Mueller for his crucial help in the determination of
the analytical form of the low-multiplicity tail of the dipole distribution.
This work would not have been possible without the intensive use of
a cluster: Therefore, we  are  grateful  to  the  CPHT computer support team, and especially
to Jean-Luc Bellon and Danh Kim Pham, for their invaluable help.
}

%%%%%%%%%%%%%%%%%%%%%%%%%%%%%%%%%%%%%%%%%%%%%%%%%%%%%%%%%%%%%%%%%%%%%%%%%%%%%%%%%%%%%%%%%%%%%%%%
%%%%%%%%%%%%%%%%%%%%%%%%%%%%%%%%%%%%%%%%%%%%%%%%%%%%%%%%%%%%%%%%%%%%%%%%%%%%%%%%%%%%%%%%%%%%%%%%

\appendix

\section{\label{sec:MC}Implementation of the modified dipole model}

In this appendix, for completeness, we describe
our numerical implementation of the color dipole model. It actually follows closely
the one in Ref.~\cite{Salam:1996nb}.
Other versions of the code have been written,
incorporating different variations of the model to make it more suitable for
phenomenological studies, such as
energy conservation or gluon recombination; see e.g. Ref.~\cite{Avsar:2005iz}.

The dipole model is a $1\rightarrow 2$ branching process. Each dipole of a given set
may split, independently of the other dipoles,
into two dipoles when the
rapidity is increased by $dy$.
For definiteness, let us consider a generic dipole whose endpoints
are labeled by the two-dimensional vectors $\uvec{x}_0$ and $\uvec{x}_1$.
The probability $dp_0$
it emits a gluon at position $\uvec{x}_2$ up to $d^2\uvec{x}_2$
was given in Eq.~(\ref{eq:dp0}):
\be
dp_0=\bar\alpha dy \frac{d^2\uvec{x}_2}{2\pi}\frac {{x}_{01}^2}{{x}_{02}^2 {x}_{12}^2}
\Theta(x_{02},x_{12};\Rir).
\tag{\ref{eq:dp0}}
\ee
This probability diverges when $\uvec{x}_2$ coincides with the endpoints
$\uvec{x}_0$ or $\uvec{x}_1$:
as well-known, the probability to split into very small dipoles can be arbitrarily large
due to the collinear singularity.
We thus need to introduce a lower cutoff $\ruv$ on the sizes of the produced dipoles
in order to get a meaningful distribution. We choose to enforce it
as sharp Heaviside $\theta$ function:
\be
dp_0\longrightarrow dp_0\times\theta(x_{02}-\ruv)
\theta(x_{12}-\ruv),
\ee
where $\ruv$ is an arbitrary ultraviolet regulator, that needs to be taken much
smaller than all distance scales relevant to our problem in such a way that it
does not affect the physical results. The value of $\ruv$ we chose in practice is checked to
satisfy this requirement in Appendix~\ref{sec:testUVcutoff}.

Having the expression of the probability that the dipole splits in the infinitesimal
rapidity interval $dy$,
we can easily write the expression of the probability that the dipole
splits (for the first time)
at finite rapidity $y$ (up to $dy$) by emitting a gluon at position $\uvec{x}_2$ (up
to $d^2\uvec{x}_2$):
\be
dp=dp_0\,\theta(x_{02}-\ruv)
\theta(x_{12}-\ruv)\,
e^{-\bar\alpha\lambda(x_{01};\ruv,\Rir)y}
\label{eq:dp}
\ee
where $\bar\alpha\lambda$ is the inverse ``lifetime'' of the dipole, namely
the inverse of the typical rapidity  interval 
between two successive dipole splittings:
\be
\lambda(x_{01};\ruv,\Rir)=
\int\frac{1}{\bar\alpha}\frac{dp_0}{dy}=
\int \frac{d^2\uvec{x}_2}{2\pi}\frac {x_{01}^2}{x_{02}^2 x_{12}^2}\theta(x_{02}-\ruv)
\theta(x_{12}-\ruv)\Theta(x_{02},x_{12};\Rir).
\ee

We use two elementary techniques to implement dipole splitting in a Monte Carlo code.
The first one is based on the following mathematics:
If $f(x)$ is the probability density of the real variable $x$, if $F(x)=\int_{-\infty}^x dx'\,f(x')$ is its
cumulative distribution function,
and if $y$ is distributed uniformly between~0 and~1, then $x=F^{-1}(y)$ is distributed according to $f$.
The algorithm that follows from this observation is practical whenever $F$ and its inverse have analytical expressions.
When this is not the case, then we use a rejection algorithm: We pick a density $\tilde f(x)$ whose
inverse cumulative
distribution may be expressed by a simple analytical formula,
and which is such that $\tilde f(x)\geq f(x)$ for all $x$.
We then generate realizations of
$x$ according to $\tilde f$, and accept the generated values with probability  $f(x)/\tilde f(x)$.

\subsection{Dipole evolution without an infrared cutoff}

We first address dipole evolution defined by the probability $dp$ in which
the infrared cutoff is put to $\Rir=+\infty$, namely with the cutoff function set to
$\Theta=1$: we shall denote this probability by $dp|_\text{pQCD}$.
We start by explaining how to generate the distribution of the position of the gluon
(or, equivalently, of the size vectors of the produced dipoles in the splitting).
Then, we generate the rapidity at which the splitting occurs.

\paragraph{Distribution of the position of the emitted gluon.}

In practice, we implement the emission of a gluon off a dipole
whose endpoints $(\uvec{x}_0,\uvec{x}_1)$
are located at positions $\uvec{x}_0=(0,0)$ and $\uvec{x}_1=(1,0)$
in the two-dimensional plane,
and use the invariance of the emission kernel $dp_0|_\text{pQCD}$
under M\"obius transformations (including
translations, rotations and dilations) in order to convert it
into an emission off a generic dipole.
Hence we shall rescale the ultraviolet
cutoff, defining ${\bar r}_\text{\tiny UV}=\ruv/x_{01}$.% and $R=\Rir/x_{01}$.
Thanks to the symmetries of dipole splitting under reflections,
we can restrict ourselves to generate gluons in the upper left quadrant
of the transverse plane, defined by
\be
x_{02}\cos\varphi<\frac12\quad\text{and}\quad
\varphi_\text{min}<\varphi<\pi
\quad\text{with}\quad
\varphi_\text{min}=\theta({\bar r}_\text{\tiny UV}-1/2)\arccos\frac{1}{2{\bar r}_\text{\tiny UV}},
\ee
and perform mirror symmetries with respect to the $(Ox)$ and $(Oy)$ axis,
with probability $\frac12$ each, in order to recover
the correct distribution in the whole transverse plane.
 
When a gluon is emitted by such a dipole, it is found at position $\uvec{x}_2$
that we shall label by the polar coordinates $(x_{02},\varphi)$,
up to $(dx_{02},d\varphi)$, or at one of the 3 positions deduced from $\uvec{x}_2$
by mirror symmetries, with the probability
  \be
  \frac{4}{\lambda(1;{\bar r}_\text{\tiny UV},\infty)}\frac{1}{\bar\alpha}\frac{dp_0|_\text{pQCD}}{dy}
  =dx_{02}\frac{d\varphi}{2\pi}\frac{4}{\lambda}\frac{\theta(x_{02}-{\bar r}_\text{\tiny UV})}{x_{02}(1+x_{02}^2-2x_{02}\cos\varphi)}
  \label{eq:probadensity}
  \ee
  where the restriction to the upper left quadrant is understood.
  (Note that the ultraviolet cutoff ${\bar r}_\text{\tiny UV}$ is fixed, so that $\lambda$ is a constant.)

The probability density $f$ of $x_{02}$ is easily determined by marginalizing the
joint density of $x_{02}$ and $\varphi$, that can be read off Eq.~(\ref{eq:probadensity}),
over the angle $\varphi$.
The integral of the probability density in Eq.~(\ref{eq:probadensity}) with respect to the angle
over the interval $[\varphi,\pi]$ reads
\be
\begin{split}
\int \frac{4}{\lambda}\frac{1}{\bar\alpha}\frac{dp_0}{dy}
&=\frac{dx_{02}}{x_{02}}\int_\varphi^\pi \frac{d\varphi^\prime}{2\pi}\frac{4}{\lambda}
\frac{\theta(x_{02}-{\bar r}_\text{\tiny UV})}{1+x_{02}^2-2x_{02}\cos\varphi^\prime}\\
&={dx_{02}}\frac{4}{\lambda}\frac{\theta(x_{02}-{\bar r}_\text{\tiny UV})}{\pi x_{02}|1-x_{02}^2|}
\arctan\left(\frac{|1-x_{02}|}{1+x_{02}}\cotan\frac{\varphi}{2}\right).
\end{split}
\label{eq:varphi}
\ee
The distribution $f$ of $x_{02}$ follows by setting $\varphi=\varphi_\text{min}$ in the previous equation:
\be
f(x_{02})=\frac{4}{\lambda}\times
\begin{cases}  
  \frac{\theta(x_{02}-{\bar r}_\text{\tiny UV})}{2 x_{02}(1-x_{02}^2)}&\quad \text{for}\quad x_{02}\leq\frac12\\
  \frac{1}{\pi x_{02}|1-x_{02}^2|}\arctan\left(\frac{|1-x_{02}|}{1+x_{02}}\sqrt{\frac{x_{02}+\frac12}{x_{02}-\frac12}}\right)
    &\quad \text{for}\quad x_{02}>\frac12.
\end{cases}
\ee
Since it is not possible to integrate analytically the function $f$, we
use a density $\tilde f(x_{02})$ to generate $x_{02}$, such that $\tilde f(x_{02})\geq f(x_{02})$,
and which admits an integral straightforward to invert analytically.
We eventually correct the distribution of $x_{02}$
with the help of a rejection algorithm. In practice, we use
\be
\tilde f(x_{02})=\frac{4}{\lambda}\frac{5}{\left[6x_{02}(1+x_{02}^2)\right]}
\ee
to generate realizations of $x_{02}$,
and accept the obtained values with probability $f(x_{02})/\tilde f(x_{02})$.

Once $x_{02}$ has been determined, $\varphi$ is easily generated since its cumulative distribution function,
deduced from Eq.~(\ref{eq:varphi}),
admits a simple inverse function.

\paragraph{Distribution of the rapidity of the first emission of one dipole.}

We also need to generate the rapidity at which this emission occurs.

Due to the independence of the dipole splittings, justified by the large-number-of-color limit,
the rapidity interval $\Delta y$ before the next emission of a gluon
off the considered dipole just follows an exponential law. Its distribution reads
\be
p(\Delta y)=\bar\alpha\lambda\, e^{-\bar\alpha\lambda\, \Delta y}.
\ee
Thus this law is completely determined by $\lambda$.
Once this parameter has been computed, realizations of $\Delta y$ are
trivial to generate.
$\lambda$ is given by a two-dimensional integral, over~$x_{02}$ and~$\varphi$:
\be
\begin{split}
\lambda(1;{\bar r}_\text{\tiny UV},\infty)&=\int_{{\bar r}_\text{\tiny UV}}^{+\infty}
\frac{dx_{02}}{2\pi x_{02}}\int_0^{2\pi}\frac{d\varphi}{1+x_{02}^{2}-2x_{02}\cos\varphi}\\
&=4\int_{{\bar r}_\text{\tiny UV}}^{+\infty}\frac{dx_{02}}{2\pi x_{02}}\int_{\theta(x_{02}-\frac12)\arccos\frac{1}{2x_{02}}}^\pi
\frac{d\varphi}{1+x_{02}^{2}-2x_{02}\cos\varphi}.
\end{split}
\ee
One can perform analytically the integral over the angle,
which leaves us with a one-dimensional integral
\begin{multline}
\lambda=\theta({\bar r}_\text{\tiny UV}-1/2)\ln\left[\frac13\left(\frac{1}{{\bar r}_\text{\tiny UV}^2}-1\right)\right]\\
+\frac{4}{\pi}\int_{\max(1/2,{\bar r}_\text{\tiny UV})}^{+\infty}\frac{dx_{02}}{x_{02}}\frac{1}{|1-x_{02}^{2}|}
\arctan
\left(
\frac{|1-x_{02}|}{1+x_{02}}\sqrt{\frac{x_{02}+\frac12}{x_{02}-\frac12}}
\right),
\end{multline}
which needs to be performed numerically.

\paragraph{Generation of a full Fock state of an onium at rapidity~$y$.}

Once one gluon emission, corresponding to a $1\rightarrow 2$ dipole
splitting, is implemented, the full Fock state at a given rapidity $y$ is
generated through a simple iteration.

Starting from one dipole $(\uvec{x}_0,\uvec{x}_1)$,
the rapidity $\Delta y$ at which it emits a gluon is generated. If this rapidity
is found to be larger than the final rapidity $y$,
then the evolution stops: the Fock state in this particular
event consists in a single dipole.
If instead it is less than~$y$, then the position $\uvec{x}_2$ of the gluon is generated (see above),
and the initial dipole is replaced by two dipoles, $(\uvec{x}_{0},\uvec{x}_2)$ and $(\uvec{x}_{2},\uvec{x}_1)$.
This procedure is then just applied recursively to the two new dipoles over the rapidity
interval $y-\Delta y$.

%%%%%%%%%%%%%%%%%%

\subsection{Enforcing the infrared cutoff}

When the cutoff $\Rir$ is set to a finite value,
i.e. when a function $\Theta\leq 1$ is added to the splitting probability, then
the integral over the angle $\varphi$ can in general not be done analytically.

Two nested numerical integrals have now to be performed to compute
the inverse lifetime $\lambda$. In practice, it proves useful to
construct a lookup table containing the numerical evaluations of
the two-dimensional integral for a set of sizes $x_{01}$.

As for the distribution of the position of the emitted gluon,
the simplest is to generate the dipole sizes with the weight
given by the dipole model without the cutoff, and
then use a rejection algorithm to correct the distribution:
The splitting of a dipole of size $x_{01}$ into two dipoles of respective sizes $x_{02}$ and
$x_{12}$ is accepted with probability $\Theta(x_{02},x_{12};\Rir)$.
Since the cutoff function $\Theta$ cuts out a low-probability region,
in which two dipoles larger than the initial one are produced,
the efficiency of such an algorithm is quite high.

\section{\label{sec:testUVcutoff}Numerical test of the effect of the UV cutoff}

\begin{figure}[ht]
  \begin{center}
  \includegraphics[width=0.7\textwidth]{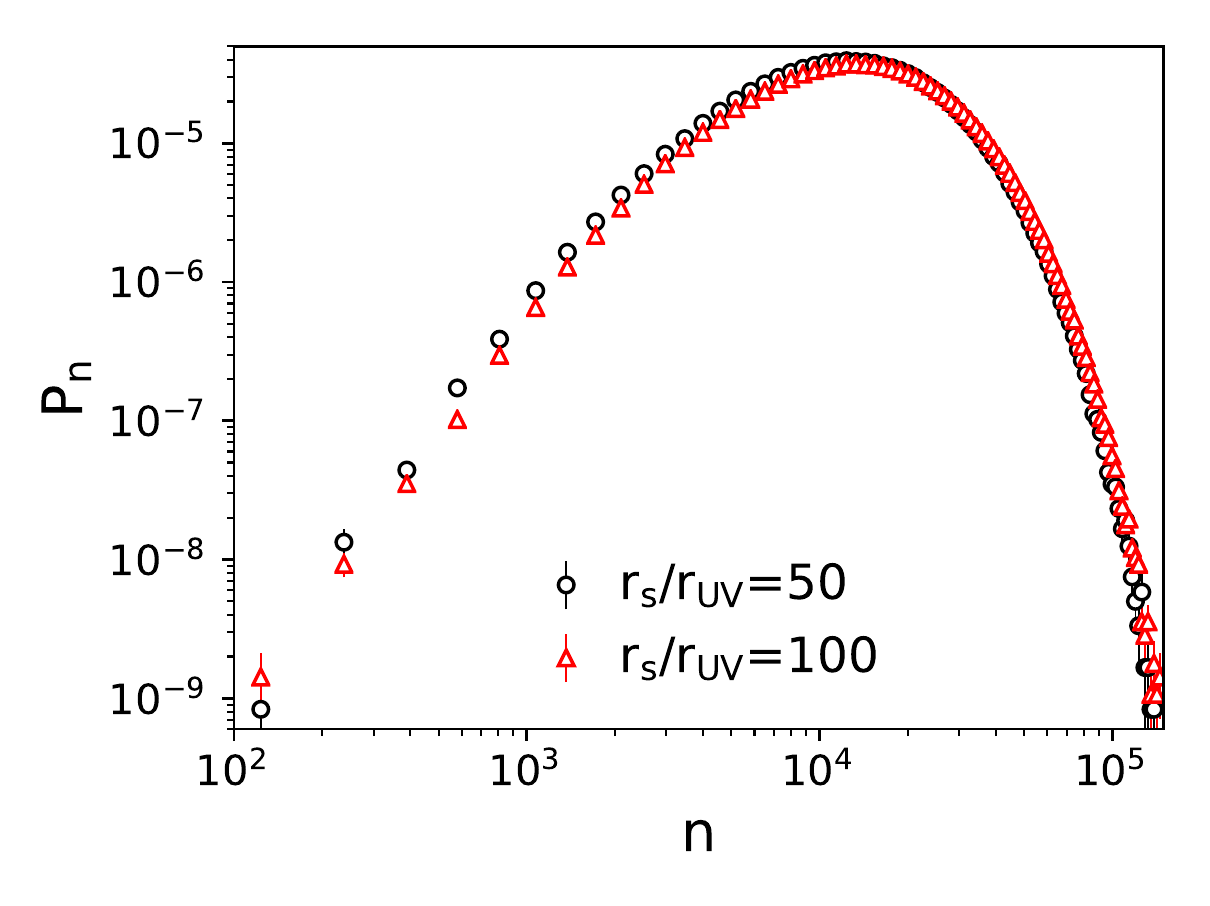}
  \end{center}
  \caption{\label{fig:effect_UV_cutoff}\small
    {\it Effect of the UV cutoff on the distribution of the multiplicity.}
    We fix the values of the saturation scale $r_s$ and of the onium size $x_{01}$,
    and pick two different values for the UV cutoff, see the legend. 
  The rapidity is set to $\bar\alpha y=4$.}
\end{figure}

The ultraviolet cutoff protecting us from the collinear divergence of dipole emission
is unphysical, and as such, any conclusion obtained within our model must be strictly
independent of its value.
This means that we need to choose a value of $\ruv$ which is small enough to yield no
effect on the shape of $P_n$.
At the same time, since the typical number of dipoles grows with $\ruv$ as a power,
we can not pick a too small value, for the sake of saving computation time.

In Fig.~\ref{fig:effect_UV_cutoff}, we display how the multiplicity distribution gets
altered if we vary the UV cutoff by a factor 2 , all other parameters being fixed: $\bar\alpha y=4$,
$r_s/\Rir=0.025$ and $x_{01}/\Rir=0.5$. 
We find that the shape is unaffected by this choice. 
Simply, as could be expected, the number of dipoles is globally slightly lower for larger cutoffs. 
For the practical calculations, we have always set $\ruv=r_s/100$, which,
we have tested, provides stable shapes of $P_n$ at $\bar \alpha y=4$.

%%%%%%%%%%%%%%%%%%%%%%%%%%%%%%%%%%%%%%%%%%%%%%%%%%%%%%%%%%%%%%%%%

\end{document}